\documentclass[aps,prb,floatfix,amsmath,amssymb,preprint,eqsecnum]{revtex4}
\usepackage{graphicx}% Include figure files
\usepackage{dcolumn}% Align table columns on decimal point
\usepackage{bm}% bold math
\usepackage{epstopdf}
\usepackage{setspace}   % controllabel line spacing
\usepackage{amsmath}
\usepackage{amssymb}

\begin{document}
\title{Non-perturbative Quantum Dynamics of the Order Parameter in the Pairing Model}

\author{Victor Galitski}
\affiliation{
Joint Quantum Institute and Department of Physics,
University of Maryland, College Park, MD 20742-4111}

\begin{abstract}
We consider quantum dynamics of the order parameter in the discrete pairing model (Richardson model) in thermodynamic equilibrium. The integrable  Richardson Hamiltonian is represented as a direct sum of Hamiltonians acting in different Hilbert spaces
of single-particle and paired/empty states. This allows us to factorize the full thermodynamic partition function into
a combination of simple terms associated with real spins on singly-occupied states and the partition function of
the quantum $XY$-model for Anderson pseudospins associated with the paired/empty states.
Using coherent-state path-integral, we calculate the effects of superconducting phase fluctuations exactly. The contribution of superconducting amplitude fluctuations to the partition function in the broken-symmetry phase is shown to follow from the  Bogoliubov-de Gennes
equations in imaginary time. These equations in turn allow several interesting mappings, {\em e.g.,}
they are shown to be in a one-to-one correspondence with the one-dimensional
Schr{\"o}dinger equation in supersymmetric Quantum Mechanics. However,  the most practically
useful approach to calculate functional determinants is found to be via an analytical continuation of the quantum
 order parameter to real time, $\Delta(\tau \to it)$, such that the problem maps onto that of a driven two-level system.  The contribution of a particular dynamic order parameter, $\Delta(\tau)$, to the partition function is shown to correspond to the sum of the Berry phase and dynamic phase accumulated by the pseudospin. We also examine a family of exact solutions for two-level-system dynamics on  a class of elliptic functions and  suggest a compact expression to estimate the functional determinants on such trajectories. The possibility of having  quantum soliton solutions co-existing with classical BCS mean-field is discussed.
\end{abstract}

\maketitle

\section{Introduction}
\label{sec:Intro}

The concept of spontaneous symmetry breaking is one of the cornerstones of modern physics:  Most phase transitions we know are associated with
the appearance of a non-zero local order parameter that represents a broken symmetry and leads to a state that has a lower symmetry
than that of the underlying Hamiltonian. In elementary particle physics, the Anderson-Higgs mechanism is the most promising scenario
to explain the appearance of finite masses for elementary particles, including gauge bosons. The canonical model to explain the
origin of the broken symmetry phenomenon usually involves a Lagrangian for a boson field, $\Delta$, that has quadratic and quartic
terms and that can be symbolically represented as follows: $L[\Delta] = \alpha \left| \Delta \right|^2 + \beta \left| \Delta \right|^4 + c  \left|D \Delta\right|^2$, where $D$ corresponds to a gauge-invariant derivative and $\alpha$, $\beta$, and $c$ are constants. In the context of elementary particle physics, it defines  a Mexican-hat model for the Higgs boson, which is a minimal renormalizable field theory that produces symmetry breaking ``by design.'' In solid state physics, such a Lagrangian is associated with the Ginzburg-Landau functional for a fluctuating order parameter near a phase transition and in many cases it can actually be derived from a more general microscopic Hamiltonian (which is typically an interacting fermion model, such that the order-parameter field is associated with a composite, rather than canonical boson). 

Such a microscopic derivation was first accomplished by Gor'kov,~\cite{Gorkov} who starting from the BCS Hamiltonian obtained the Ginzburg-Landau functional for a superconductor and found explicitly the Ginzburg-Landau coefficients  in terms of microscopic parameters ({\em i.e.}, electron mass, electron density, interaction strength, and concentration of impurities). The general framework for a derivation of this type now appears in excellent textbooks~\cite{ASbook} and can be briefly summarized as follows: One starts with an interacting electron model that has a ``desired'' phase transition ({\em e.g.}, electrons with attraction for superconductivity): The partition function of the model can be expressed  in terms a path integral of the corresponding imaginary-time (Grassmann) action, which includes a quartic term describing interactions. This term in path integral can be decoupled via an auxiliary Hubbard-Stratonovich boson field, $\Delta(x) \equiv \Delta(\tau,{\bf r})$.
Then, the fermionic component of the action becomes Gaussian and the fermions can be integrated out to produce an effective action $S_{\rm eff} [\Delta(x)]$, which can be formally expressed as a non-linear functional determinant [see, {\em e.g.}, Eq.~(\ref{Seff}) in Sec.~\ref{sec:HS}]. The Hubbard-Stratonovich field, $\Delta(x)$, describes a fluctuating in space and imaginary time, $\tau$, order parameter and the appearance of a non-zero expectation value, $\overline{\Delta}$, of this field below a certain transition temperature, $T_{\rm c}$, is associated with a broken symmetry phase. In the vicinity of $T_{\rm c}$, the relevant trajectories of $\Delta(x)$ are assumed to be such that its imaginary-time dependence is unimportant [that is, $\Delta(\tau,{\bf r})$ is assumed to be independent of $\tau$], $\Delta({\bf r})$ is in some sense small, and it is also assumed to be weakly fluctuating in space (long-wavelength approximation). Hence, the action can be related to the free energy by simply writing ${\cal F}\left[ \Delta({\bf r}) \right] = T_{\rm c} S\left[ \Delta({\bf r}) \right]$, and  expanded in a Taylor series, which yields the Ginzburg-Landau theory, with the quadratic coefficient $\alpha \propto \left( T - T_{\rm c} \right)$.

The derivation of the Ginzburg-Landau theory outlined above is justified only near a classical phase transition. Below $T_{\rm c}$, the assumptions about $\Delta$ being small and $\tau$-independent break down (if the relevant interaction constant, $g$, is not small they may break down even ``earlier''). However, it is exactly the low-temperature phase, including the ground state that we associate with a spontaneously broken symmetry. This picture is based on the very reasonable assumption that the relevant ``trajectories'' of the order-parameter field, $\Delta(x)$, at low temperatures are located near the classical saddle-point $\overline{\Delta} \equiv {\rm const}$, which becomes the only possible trajectory at $T = 0$ and therefore represents an exact solution. This assumption is equivalent to stating that the effective action, $S_{\rm eff} \left[ \Delta(x) \right]$, has one and only one minimum which occurs in a single ``point'' in the space of all allowed functions, $\left| \Delta(\tau,{\bf r})\right|$ (modulo the overall phase). We reiterate that there is no good reason to expect that the simplified form of the Ginzburg-Landau action remains reliable at low temperatures. In fact, if we ``insist'' on the canonical Ginzburg-Landau form and attempt to derive the corresponding coefficients in the expansion, we shall find that the coefficient  of the quartic term generally diverges as $T \to 0$.~\cite{VG1,VG2} Hence, we have to work with the full functional determinant in $S_{\rm eff} \left[ \Delta(x) \right]$, which is a  complicated non-linear functional and we know little about its properties apart from its behavior in a tiny sub-space of constant functions. To the best of author's knowledge, there is no  model (associated with breaking of continuous symmetry below $T_{\rm c} >0$), where such  functional determinants have been explicitly calculated beyond the classical mean-field analysis.

The objectives of this work are to bring up the general problem of non-perturbative quantum dynamics in broken-symmetry phases and to construct a general framework to calculate functional determinants that appear in the {\em non-linear} effective action for {\em quantum}  trajectories of the order parameter in the pairing model. The latter is a seemingly hopeless goal, but we show that one can obtain exact results in certain cases and based on those results formulate a more general Ans{\"a}tz that is expected to be useful for a large class of quantum trajectories. To address this and other related questions, we employ the Richardson pairing model,~\cite{Rich,Rev1,Rev2} which is an interacting fermion model that has a paired ground state built-in. In fact, it is ``almost'' the mean-field BCS model in the sense that the corresponding order parameter does not have any real-space dependence and so all such fluctuations~\cite{AL} have been eliminated. However, the model still retains quantum dynamics of $\Delta(\tau)$.  The Richardson model is integrable and there exists an exact Bethe-Ans{\"a}tz solution,~\cite{Rich,Rev2} which determines the exact eigenstates and spectrum of the model in sectors with a fixed number of single-particle excitations and Cooper pairs. However,  this algebraic Bethe Ans{\"a}tz solution does not appear to be very helpful in calculating the thermodynamic partition function in the grand-canonical ensemble and we use an alternative method, which is based on coherent-state path-integral representation of Anderson pseudospins,~\cite{PA} describing the BCS sector of the model.  We use a mapping of the equilibrium problem in imaginary time onto that of non-equilibrium superconductivity, and take advantage of the exact non-equilibrium solutions, obtained recently in a series of amazing papers by Levitov {\em et al.}~\cite{BLS} and Yuzbashyan {\em et al.}~\cite{EY0,EY1} By analyzing a certain family of exact results, we propose a general closed expression to estimate the corresponding functional determinant, which is not always exact but is expected to be quantitatively reliable for a large class of elliptic functions and their limits.

Our paper is structured as follows: In Sec.~\ref{sec:HS}, we present the canonical Richardson model and formulate in more technical details the key questions within the conventional Grassmann path integral/Hubbard-Stratonovich approach. The questions involve studying various aspects of fluctuation physics and they are addressed in the rest of the manuscript using a variety of techniques: In Sec.~\ref{sec:GRH}, we derive combinatorially an exact expression for the thermodynamic partition function of a generalized Richardson model in terms of a ``spin partition function'' associated with single-particle states and  an ``Anderson pseudospin partition function'' associated with the paired/empty states. The generalized Richardson model includes the canonical Richardson model (reduced BCS Hamiltonian) as a particular case, and in this limit, the spin part of the partition function becomes trivial, so that the problem reduces to the problem of calculating contributions of Anderson pseudospins to the partition function. Sec.~\ref{sec:CSPI} formulates a coherent-state path integral for Anderson pseudospins to calculate the functional determinants of interest. It is shown that by introducing a single Hubbard-Stratonovich field one can represent
the full thermodynamic partition function as a product of terms local in parameter space. The contribution of each such local term to the partition function follows from the Bogoliubov-de Gennes equation in imaginary time. In Sec.~\ref{sec:phase}, we study phase fluctuations within the path integral formalism and obtain an exact expression for the partition function in terms of a sum of phase winding numbers. Sec.~\ref{sec:amplitude} is the main part of the paper, which addresses the question of (possible) fluctuations of the amplitude of the order parameter, assuming that the phase fluctuations are completely suppressed. Sec.~\ref{sec:amplitude} contains several parts: In Sec.~\ref{sec:BdG}, the symmetry properties of the imaginary-time Bogoliubov-de~Gennes
equations are discussed and it is shown that the full density matrix solution satisfying the proper initial condition, $\rho(\tau\to 0) = \hat{1}$, can be constructed from a particular spinor  solution satisfying arbitrary initial conditions. In Sec.~\ref{sec:SUSY}, we show that the general problem of solving  imaginary-time Bogoliubov-de~Gennes equations in the presence of a quantum-fluctuating order-parameter field  is equivalent to that of a one-dimensional supersymmetric Schr{\"o}dinger equation, with ``superpotentials'' determined uniquely by $\Delta(\tau)$. Therefore, the cases where these two problems are solvable are shown to be closely related. Sec.~\ref{sec:Solvable} derives an exact expression for the full density matrix, $\hat{\rho}(\tau)$, corresponding to
a non-trivial dynamic order parameter, representing the soliton of Ref.~[\onlinecite{BLS}] analytically-continued to imaginary time. The resulting functional determinant is found to be surprisingly simple and is equivalent to that of a Fermi gas. Sec.~\ref{sec:FDet} suggests that the simplification of the functional determinant observed in Sec.~\ref{sec:Solvable} is not accidental but has a natural explanation: It is argued that the effective action associated with a given quantum-fluctuating, $\Delta(\tau)$, is given by the sum of the dynamical phase and Berry phase accumulated by a two-level-system driven by a time-dependent magnetic field determined by the analytically continued order parameter, $\Delta(\tau \to it)$.
This conjecture is verified to work well on a large class of functions, where $\Delta(\tau + it)$ is an elliptic function with two primitive periods along the $\tau$ and $it$-axes. A general expression for the corresponding effective action is presented in Sec.~\ref{sec:Seff} and the possible implications of the results obtained to non-perturbative quantum dynamics of the superconducting order parameter are discussed.

\section{The Richardson Pairing Model and Key Questions}
\label{sec:HS}

 Let us consider spin-$1/2$ fermions, described by the creation/annihilation operators, $\hat{c}^\dagger_{l,s}$ and $\hat{c}_{l,s}$,
labeled by the spin index $s =\pm1$ and the index $l \in {\cal L}$, where ${\cal L}$ is a set of allowed single-particle states. It can be a discrete, possibly finite, set (associated for example with localized levels in a mesoscopic superconducting grain~\cite{R21,R22,Ioffe_etal}) or a continuum of momentum states in a system with open boundary conditions (such that $\left| l, s \right\rangle$ and $\left| l, -s \right\rangle$ are a pair of time-reversed states). We will refer to the states $ l $ as to ``sites.'' We perform some formal mathematical manipulations assuming that ${\cal L}$ is discrete and finite, but it is without a loss of generality, as this assumption does not preclude us from taking the proper limit at any stage of the calculation. The canonical Richardson Hamiltonian (or equivalently the reduced BCS Hamiltonian) describing an $s$-wave superconductor has the form:
\begin{equation}
\label{CRH}
\hat{\cal H}_{\rm R} = \sum\limits_{l \in {\cal L}, s =\pm} \xi_l \hat{c}^\dagger_{l,s}  \hat{c}_{l,s}
- {g \over 2 V_{\cal L}} \sum\limits_{l,l' \in {\cal L}} \hat{c}^\dagger_{l,+}  \hat{c}^\dagger_{l,-}  \hat{c}_{l',-}  \hat{c}_{l',+},
\end{equation}
where $V_{\cal L}$ is either the number of sites in ${\cal L}$, if the set ${\cal L}$ is discrete, or otherwise if  ${\cal L}$ represents  a continuum spectrum, $V_{\cal L}$ is  a volume (in this case, the sums are to be replaced with integrals over momenta, ${\bf k} \equiv l$).
 In what follows, we will also use the notation $\tilde{g} = g/\left( 2 V_{\cal L} \right)$.

To formulate the main questions, let us first follow the conventional method of treating Hamiltonian (\ref{CRH}) and represent the partition function  as a Grassmann path integral:
\begin{equation}
\label{ZGpi}
Z_R = \int {\cal D} \bar{c}_{l,s}[\tau] {\cal D} {c}_{l,s}[\tau] \exp\left\{ - \int\limits_0^\beta d\tau \sum\limits_{l, l' \in {\cal L}}
\Bigl[ \delta_{l,l'} \bar{c}_{l,s}  \left( \xi_l - \partial_\tau\right) c_{l,s} - \tilde{g} \bar{c}_{l,+} \bar{c}_{l,-} {c}_{l',-} {c}_{l',+} \Bigr] \right\}.
\end{equation}
 Now, we use the identity, $e^{\tilde{g}\, \overline{c}\, \overline{c}' c'\, c} = \int {d^2 \Delta \over \pi |\tilde{g}|}
  \exp{\left[ - {1 \over \tilde{g}} \left| \Delta \right|^2 + {1 \over \sqrt{2}} \left(\Delta\overline{c}\, \overline{c}' + \overline{\Delta} c' c \right) \right]}$, and introduce the Hubbard-Stratonovich field, $\Delta(\tau)$, to decouple the interaction term in the Grassmann action, integrate out the fermions from the resulting quadratic theory, and arrive to the following standard effective action expressed in terms of the order parameter field
\begin{equation}
\label{ZGHS1}
Z_R = \int {\cal D} \Delta^*[\tau] {\cal D} \Delta[\tau] e^{-{1 \over \tilde{g}} \int\limits_0^\beta d\tau \left| \Delta(\tau) \right|^2 }
\prod_{l \in {\cal L}} {\rm  Det}\, \left[ \partial_\tau - \xi_l\, \hat{\tau}^z - {\rm Re}\, \Delta\, \hat{\tau}^x + {\rm Im}\, \Delta\,  \hat{\tau}^y \right],
\end{equation}
where $\hat{\bm \tau}$ are two-by-two Pauli matrices in the Nambu space and the determinant is to be evaluated over both the time variable and the Nambu space. To trace over the Nambu space, one can use the identity below
\begin{equation}
\label{Mat_identity}
{\rm  Det}\, \Biggl| \Biggl|
\begin{array}{cc}
A & B\\
C & D
\end{array}
\Biggr| \Biggr| = {\rm det}\, \left[ A \cdot D \right] {\rm det}\, \left[ 1 - D^{-1} \cdot C \cdot A^{-1}\cdot B \right],
\end{equation}
which is valid for any matrices/operators $A$, $B$, $C$, and $D$, provided that $A$ and $D$ are invertible. Applying this identity to Eq.~(\ref{ZGHS1}), we find
\begin{equation}
\label{ZGHS11}
Z_R = Z_{\rm  FG}  \int {\cal D} \Delta^*[\tau] {\cal D} \Delta[\tau]
\exp{\left\{-{1 \over \tilde{g}} \int\limits_0^\beta d\tau \left| \Delta(\tau) \right|^2 + S_{\rm eff} \left[ \Delta(\tau) \right] \right\}},
\end{equation}
where the effective action is given by
\begin{equation}
\label{Seff}
 S_{\rm eff} \left[ \Delta(\tau) \right] = \sum\limits_{l \in {\cal L}}
{\rm Tr}\, \ln \left[ 1 - G^+_l \cdot \Delta^* \cdot G^-_l \cdot \Delta \right]
\end{equation}
and $Z_{\rm  FG} = \prod_{l \in {\cal L}} {\rm det}\, \left[ \partial_\tau^2 - \xi_l^2 \right]$ is the partition function of a non-interacting Fermi
gas given by Eq.~(\ref{ZFG}) below. Here $G^\pm_l = \left( \partial_\tau \pm \xi_l\right)^{-1}$ are Green functions, whose explicit form in $\tau$-representation
is easy to obtain.

Calculating formally the first variation of the effective action $S_{\rm eff} \left[ \Delta \right]$ with respect to  $\Delta[\tau]$ leads to the mean-field equation
for an extremum $\Delta_{\rm MF} [\tau]$ of the functional, which generally has the complicated operator form:
\begin{equation}
\label{MFSP}
{1\over \tilde{g}} \Delta_{\rm MF}^*(\tau) \delta(\tau - \tau') = \sum\limits_{l \in {\cal L}} \left[ 1 - G^+_l \cdot \Delta_{\rm MF}^* \cdot G^-_l \cdot \Delta_{\rm MF} \right]^{-1}
\cdot \left[ G^+_l  \cdot \Delta_{\rm MF}^* \cdot G^-_l \right](\tau,\tau'),
\end{equation}
where the right-hand-side is to be understood as a kernel of the corresponding operator in $\tau$-representation $\Bigl[$ {\em i.e.}, a kernel, $K(\tau,\tau')$, defines an operator by its action on an arbitrary $\beta$-periodic function, $f(\tau)$, as follows $ \hat{K} \cdot f(\tau) = \int_0^\beta K(\tau,\tau') f(\tau') d\tau'$ $\Bigr]$. The equation (\ref{MFSP}) can be cast into a more
friendly form of an integral equation, but it would still remain too complicated for a systematic analysis. We do however know that there exists a solution to this equation, which is a constant that in the classical BCS model is given by  $\overline{\Delta}_{\rm BCS} \sim \tilde{\omega}_0 e^{-1/(\nu g)}$ (here we have to assume that ${\cal L}$ is momentum space and $\sum\limits_{l \in {\cal L}} \cdot = V_{\cal L} \nu \int  d \xi_l \cdot$, $\nu$ is the density of states at the Fermi level, and $\tilde{\omega}_0$ is the usual high-energy cut-off to regularize the Cooper logarithm). One can verify explicitly that indeed $\overline{\Delta}_{\rm BCS}$ is a true minimum [{\em i.e.}, it is not only a minimum on a tiny subset of constant functions, but a true minimum on the space of allowed functions, $\Delta(\tau) = \Delta(\tau + \beta)$], but there are still a few important questions that remain:
{\bf (i)}~Does the classical BCS mean-field result represent the only minimum at $0 \leq  T < T_{\rm c}$, or there may exist quantum non-perturbative trajectories of $\Delta(\tau)$, which would give contributions energetically comparable to the classical mean-field (or better)?
{\bf (ii)}~A related key technical question is whether it is possible to calculate the functional determinant, ${\rm det}\, \left[ 1 - G^+_l \cdot \Delta^* \cdot G^-_l \cdot \Delta \right]$, for ``trajectories'' of the order parameter with non-trivial quantum dynamics?
{\bf (iii)}~What are the effects of quantum fluctuations~\cite{AVAI,VGAIL} of the modulus and/or phase of the order parameter on thermodynamics ({\em e.g.},  the energy of the ground state)? We will address these questions to some extent in the following sections using an alternative method, namely the path-integral formalism  for Anderson pseudospins.

\section{Factorization of the Generalized Richardson Hamiltonian}
\label{sec:GRH}

The Richardson Hamiltonian (\ref{CRH}) is known to be integrable~\cite{Rich,Rev2} and its integrability is due to the existence of an infinite number of conservation laws at two levels of the problem: First, the Hamiltonian commutes with the $z$-component of the spin on any site and therefore the Hilbert sub-spaces associated with the singly-occupied states and the paired/empty states are separated and can be studied independently.~\cite{Soloviev} After this factorization, the Hamiltonian
for paired states reduces to a pseudospin  Hamiltonian (expressed in terms of Anderson pseudospins). As Richardon discovered, the pseudospin Hamiltonian amazingly has an infinite number of conservation laws as well and this allowed him to construct an exact Bethe-Ans{\"a}tz solution to the corresponding spin problem in a given sector (with a fixed total pseudospin), and in particular, find a set of coupled algebraic equations determining the energy spectrum in the sector. The Richardson equations are exact and therefore include correctly all quantum fluctuation effects, but this exactness also makes it difficult to use the solution for practical purposes and to interpret its physical meaning, because the solution mixes up fluctuations of the order parameter of different types.  In addition, the Richardson equations are still too complicated to allow a further analytic treatment and most importantly they address different pseudospin sectors independently. For this reason, we do not use the results of the algebraic Bethe-Ans{\"a}tz approach to calculate thermodynamic properties of the model, but we find it however very useful to perform the first simpler step in the Richardson's solution, {\em i.e.}, to factorize the Hilbert space into single-particle and paired/empty states. It turns out that this factorization is allowed for a more general Hamiltonian than (\ref{CRH})  and in the interest of generality and future work, we present this procedure for such a more general model, which we dub the generalized Richardson model [{\em e.g.} Eq.~(\ref{GRH}) below represents a generalized Ising-Richardson model].

Let us define the density, spin, and Cooper pair operators on each site as follows:
\begin{equation}
\label{density}
\hat{\rho}_l = \sum\limits_{s =\pm} \hat{c}^\dagger_{l,s} \hat{c}_{l,s}\,\,\, \mbox{ is the density operator},
\end{equation}
\begin{equation}
\label{spin}
\hat{\bf S}_l = {1 \over 2} \sum\limits_{s =\pm} \hat{c}^\dagger_{l,s} {\bm \sigma}_{s,s'} \hat{c}_{l,s'}
\end{equation}
is the spin, with $\hat{S}_l^z = {1 \over 2} \sum\limits_{s =\pm} s\, \hat{c}^\dagger_{l,s}  \hat{c}_{l,s}$ being its $z$-component, and
\begin{equation}
\label{CPO}
\hat{P}_l^\dagger = {1 \over 2} \sum\limits_{s =\pm}  s\, \hat{c}^\dagger_{l,s}  \hat{c}^\dagger_{l,-s}\,\,\, \mbox{ is the Cooper pair operator}.
\end{equation}
Clearly $\hat{P}_l \equiv \left( \hat{P}_l^\dagger\right)^\dagger =  \hat{c}_{l,-}  \hat{c}_{l,+}$.

Let us now use these operators to express the following generalized  Ising-Richardson model:
\begin{equation}
\label{GRH}
\hat{\cal H}_{\rm GR} = \sum\limits_{l \in {\cal L}} \left[ \xi_l \hat{\rho}_l - B_l \hat{S}_l^z \right]
- \sum\limits_{l_1,l_2 \in {\cal L}} \left[ \tilde{g}_{l_1,l_2}  \hat{P}_{l_1}^\dagger \hat{P}_{l_2} -
 \tilde{J}_{l_1,l_2}\hat{S}_{l_1}^z \hat{S}_{l_2}^z \right],
\end{equation}
where $\xi_l$ describes single-particle energy eigenvalues/spectrum, $B_l$ is an applied magnetic field in the $z$-direction,
$\tilde{g}$ is an interaction in the BCS channel, and $\tilde{J}$ is an Ising-type spin interaction. We reiterate that the special case of Hamiltonian (\ref{GRH})
with $\tilde{J} = B = 0$ and $\tilde{g}_{l_1,l_2} \equiv \tilde{g} = {\rm const}$ yields the canonical s-wave Richardson pairing model (\ref{CRH}) that we actually will study in the rest of the paper.

However, the more general Hamiltonian (\ref{GRH}) has the same ``local'' in ${\cal L}$ conservation laws,
since it commutes with the $z$-component of the spin, $\hat{S}_l^z$, on any site:
\begin{equation}
\label{loc_cl}
\left[ \hat{\cal H}_{\rm GR},\, \hat{S}_l^z \right] ={0},\, \forall l \in {\cal L}.
\end{equation}
This allows us to define the following projectors for an arbitrary subset of ${\cal L}$
\begin{equation}
\label{P1}
\hat{{\cal P}}_1 \left[ {\cal L}_1 \right] = \prod_{l \in {\cal L}_1} \left( 2 \hat{S}_l^z \right)^2
\end{equation}
and
\begin{equation}
\label{P2}
\hat{{\cal P}}_2 \left[ {\cal L}_2 \right] = \prod_{l \in {\cal L}_2}\left[\hat{1} - \left( 2 \hat{S}_l^z \right)^2\right],
\end{equation}
where ${\cal L}_{1,2} \subset {\cal L}$. Note also that $\hat{{\cal P}}_{1,2}^2 \left[ {\cal L}_{1,2} \right] = \hat{{\cal P}}_{1,2} \left[ {\cal L}_{1,2} \right]$.
By convention we shall denote the projectors on a single site ({\em i.e.}, if the corresponding subset consists of
a single element, ${\cal L}_{1,2} = \left\{ l \right\}$) as follows: $\hat{{\cal P}}_1 (l) =  \left( 2 \hat{S}_l^z \right)^2$ and
$\hat{{\cal P}}_2 (k) =\hat{1} - \left( 2 \hat{S}_l^z \right)^2$. Obviously for those single-site projectors we have:
\begin{equation}
\label{p1+p2}
\hat{{\cal P}}_1 (l) + \hat{{\cal P}}_2 (l) =\hat{1}.
\end{equation}
This resolution of unity allows us to represent the Hamiltonian (\ref{GRH}) as a sum of Hamiltonians acting in different ``sectors'' of the Hilbert space as follows:
\begin{equation}
\label{fact}
\hat{\cal H}_{\rm GR} \equiv \prod_{l \in {\cal L}} \left[ \hat{{\cal P}}_1 (l) + \hat{{\cal P}}_2 (l) \right] \hat{\cal H}_{\rm GR}=
\sum\limits_{{\cal L}_1 \cup {\cal L}_2 = {\cal L}}\hat{{\cal P}}_1 \left[ {\cal L}_1 \right] \hat{{\cal P}}_2 \left[ {\cal L}_2 \right] \hat{\cal H}_{\rm GR}.
\end{equation}
Each term in the above sum represents two Hamiltonians acting on single-particle states in ${\cal L}_1$ and paired/empty states in ${\cal L}_2$. The corresponding spin and pairing Hamiltonians are
\begin{equation}
\label{H1}
\hat{\cal H}_{\rm spin} \left[{\cal L}_1\right] = \hat{{\cal P}}_1 \left[ {\cal L}_1 \right]
\left\{ \sum\limits_{l \in {\cal L}_1} \left( \xi_l \hat{\rho}_l - B_l \hat{S}_l^z \right) +
\sum\limits_{l_1,l_2 \in {\cal L}_1} \tilde{J}_{l_1,l_2} \hat{S}_{l_1}^z  \hat{S}_{l_2}^z \right\}
\end{equation}
and
\begin{equation}
\label{H2}
\hat{\cal H}_{\rm BCS'} \left[{\cal L}_2\right] = \hat{{\cal P}}_2 \left[ {\cal L}_2 \right]
\left\{ \sum\limits_{l \in {\cal L}_2} \xi_l \hat{\rho_l} -
\sum\limits_{l_1,l_2 \in {\cal L}_1} \tilde{g}_{l_1,l_2} \hat{P}_{l_1}^\dagger  \hat{P}_{l_2} \right\}.
\end{equation}
Now, one can follow Anderson and check that the operators $\hat{P}_{l}^\dagger$,  $\hat{P}_{l}$, and $(\hat{\rho_l} - \hat{1})$,
when constrained by the projector on empty/paired states, form a closed $\mathfrak{su}(2)$ algebra on each site (here and below, we use the symbol, $\mathfrak{su}(2)$, for the Lie algebra and $SU(2)$ for the Lie group) or in other words, the
operators are Anderson pseudospins. One can therefore drop the projectors and replace the operators with Pauli matrices (since, $\hat{P}_{l}^2=0$, we have to use the two-dimensional representation) $\hat{P}_{l}^\dagger = \hat{\tau}^+_l$, $\hat{P}_{l}= \hat{\tau}^-_l$, and $\hat{\rho}_l =  \hat{\tau}_l^z + 1$.
Similarly, one can remove the projector in Eq.~(\ref{H1}) and simply replace $\hat{\rho}_l$ with one, since each site in ${\cal L}_1$ is guaranteed
to be singly-occupied by construction. This leads to the following decomposition of the Hamiltonian (\ref{GRH})
\begin{equation}
\label{decomp}
\hat{\cal H}_{\rm GR} = \sum\limits_{{\cal L}_1 \cup {\cal L}_2 = {\cal L}}
\left\{ \hat{\cal H}_{\rm  spin} \left[{\cal L}_1\right]  + \hat{\cal H}_{\rm BCS'} \left[{\cal L}_2\right] \right\},
\end{equation}
where $\hat{\cal H}_{\rm spin} \left[{\cal L}_1\right]$ and   $\hat{\cal H}_{\rm BCS'} \left[{\cal L}_2\right]$ are spin-$1/2$ Hamiltonians acting in different
Hilbert spaces. These spin Hamiltonians are of Ising and $XY$-type correspondingly:
\begin{equation}
\label{H1s}
\hat{\cal H}_{\rm spin} \left[{\cal L}_1\right] = \sum\limits_{l \in {\cal L}_1} \left( \xi_l  - {1 \over 2} B_l \hat{\sigma}_l^z \right) +
{1\over 4} \sum\limits_{l_1,l_2 \in {\cal L}_1} \tilde{J}_{l_1,l_2} \hat{\sigma}_{l_1}^z  \hat{\sigma}_{l_2}^z
\end{equation}
and
\begin{equation}
\label{H2s}
\hat{\cal H}_{\rm BCS'} \left[{\cal L}_2\right]  =
 \sum\limits_{l \in {\cal L}_2}  \xi_l  \left( \hat{\tau}^z_l + 1\right)-
\sum\limits_{l_1,l_2 \in {\cal L}_2} \tilde{g}_{l_1,l_2} \hat{\tau}^+_{l_1} \hat{\tau}^-_{l_2}.
\end{equation}
Since the Hamiltonian (\ref{GRH}) does not have operators that connect different partitions of ${\cal L}$, the total partition function
is given by a combination of the products of  the partition functions corresponding to the Ising and $XY$-models on different sets
\begin{equation}
\label{Ztot}
Z_{\rm GR} = \sum\limits_{{\cal L}_1 \cup {\cal L}_2 = {\cal L}} Z_{\rm spin}[{\cal L}_1] \times  Z_{BCS'} [{\cal L}_2].
\end{equation}
Note that  factorization of the Hilbert space into single-particle and pair/empty states, which  led us to Eq.~(\ref{Ztot}), does not  require
that $\hat{S}^z_l$ is locally conserved, but requires only that the spin and pseudospin sectors can be uncoupled via projectors (\ref{P1}) and (\ref{P2}), which is a much weaker requirement. This implies that this construction may be applied to even more general Hamiltonians of type (\ref{GRH}),  which include quantum interaction terms for real spin. This avenue will be explored elsewhere,~\cite{G} but here we instead focus on the much simpler canonical Richardson pairing Hamiltonian~(\ref{CRH}), where there are no interactions for real spins ($\tilde{J}_{l_1,l_2} \equiv 0$), nor there are magnetic fields ($B_l = 0$), and hence the partition function associated with the single particle states is simply $Z_{\rm spin} \left[ {\cal L}_1 \right] = \prod_{l\in {\cal L}_1} \left( 2 e^{-\beta \xi_l} \right)$, so that the full partition function of the pairing model is simplified to
\begin{equation}
\label{ZR}
Z_R =
\sum\limits_{{\cal L}_1 \cup {\cal L}_2 = {\cal L}} \prod_{l_1 \in {\cal L}_1} \left[ 2 e^{-\beta \xi_{l_1}} \right] Z_{\rm BCS'} [{\cal L}_2],
\end{equation}
where  $Z_{\rm BCS'} [{\cal L}_2]$ is the partition function of the $XY$-model given by $\hat{\cal H}_{\rm BCS'} \left[{\cal L}_2\right]$ in Eq.~(\ref{H2s}) on a subset ${\cal L}_2 \subset {\cal L}$ and where $\tilde{g}_{l_1,l_2} \equiv g/\left(2 V_{\cal L}\right)$. We will use this decomposition (\ref{ZR}) in the remainder of the paper.

To run a simple sanity check on the result obtained, we consider the non-interacting case with $g =0$, {\em i.e.}, the Fermi gas.
Eq.~(\ref{H2s}) therefore is the Hamiltonian of non-interacting pseudospins in magnetic fields, ${\bf b}_l = (0,0,\xi_l)$, and the partition function is given by
\begin{equation}
\label{ZFG1}
Z_{\rm FG} = \sum\limits_{{\cal L}_1 \cup {\cal L}_2 = {\cal L}} \prod_{l_1 \in {\cal L}_1} \left[ 2 e^{-\beta \xi_{l_1}} \right] \times \prod_{l_2 \in {\cal L}_2} \left[ e^{-\beta \xi_{l_2}} 2 \cosh{(\beta \xi_{l_2})} \right].
\end{equation}
Since the partition function involves products of ``local'' in ${\cal L}$ terms and all possible decompositions are to be considered, we can equivalently rewrite
Eq.~(\ref{ZFG1}) as follows
\begin{equation}
\label{ZFG}
Z_{\rm FG} =  \prod_{l \in {\cal L}} \left[ 2 e^{-\beta \xi_{l}} + e^{-\beta \xi_{l}} 2 \cosh{(\beta \xi_{l})} \right]
= \prod_{l \in {\cal L}} \left[ 1 + e^{-\beta \xi_l} \right]^2,
\end{equation}
which is indeed the partition function of a non-interacting Fermi gas of spin-$1/2$ particles.

\section{Path Integral for Anderson Pseudospins}
\label{sec:CSPI}

In Sec.~\ref{sec:GRH}, we showed that the full partition function of the Richardson model is given by
\begin{equation}
\label{ZRic}
Z_R =
 \exp\Bigl[ - \beta \sum\limits_{l \in {\cal L}} \xi_l \Bigr]  \sum\limits_{{\cal L}_1 \cup {\cal L}_2 ={\cal L}} \Bigl(  \prod_{l_1 \in {\cal L}_1} 2 \Bigr) \times
Z_{\rm BCS} [{\cal L}_2],
\end{equation}
where $Z_{\rm BCS}$ is the partition function of the $XY$-Hamiltonian with infinite-range interactions [here, we subtract a constant from the Hamiltonian $\hat{\cal H}_{\rm BCS'}$ given by Eq.~(\ref{H2s}) and set $\tilde{g}_{l_1,l_2} \equiv \tilde{g}$]:
\begin{equation}
\label{Hxy}
\hat{\cal H}_{\rm BCS} \left[{\cal L}_2\right]  =
 \sum\limits_{l \in {\cal L}_2}  \xi_l \hat{\tau}^z_l - {g \over 2 V_{\cal L}}
\left( \sum\limits_{l \in {\cal L}_2}  \hat{\tau}^+_l \right) \times \left( \sum\limits_{l' \in {\cal L}_2}  \hat{\tau}^-_{l'} \right).
\end{equation}
To calculate the partition function, we employ the coherent-state spin path integral formalism and write it in the form:
\begin{equation}
\label{Z2p}
Z_{\rm BCS} \left[{\cal L}_2\right]  = \int \left[ \prod_{l \in {\cal L}_2} {\cal D} \Omega_l (\tau) \right]
\exp \left\{ - \int\limits_0^\beta d\tau \left[ \sum\limits_{l \in {\cal L}_2} \left( -{i \over 2} \dot{\phi}_l \left[1 + \eta_l \right] + \xi_l \eta_l \right)
- \tilde{g} \sum\limits_{l,l' \in {\cal L}_2}  n^+_l  n^-_{l'}  \right]\right\},
\end{equation}
where ${\bf n}_l = \left( \sin\theta_l \cos\phi_l, \sin\theta_l \sin\phi_l, \cos\theta_l\right)$ is a vector constrained to move on a unit sphere, $\eta_l = \cos{\theta_l}$, $n^\pm_l = \left( n^x_l \pm i n^y_l\right)/2$, and $d\Omega_l(\tau_i) = d\phi_l(\tau_i) d\eta_l(\tau_i)$ for any imaginary time, $\tau_i \in \left[0,\beta=1/T\right]$.

We now perform the Hubbard-Stratonovich decoupling for the interaction term in the spin path integral, which allows us to write the full partition function
in the form:
\begin{equation}
\label{Zfull}
Z_R =
 \exp\Bigl[ - \beta \sum\limits_{l \in {\cal L}} \xi_l \Bigr]
 \int {\cal D}^2 \left[ {\Delta(\tau) \over \pi \tilde{g}} \right] e^{-{1\over \tilde{g}} \int\limits_0^\beta d\tau \left| \Delta(\tau) \right|^2 }
 \prod_{l \in {\cal L}} \left\{ 2 + z_l\left[\Delta(\tau)\right] \right\},
\end{equation}
where the $z_l$ is a {\em ``local''} path-integral, which depends on a  realization of the ``global'' Hubbard-Stratonovich field
\begin{equation}
\label{zk}
z_l\left[\Delta(\tau)\right]  =
 \int {\cal D}\Omega(\tau)  \exp \left\{ - \int\limits_0^\beta d\tau  \left[-{i \over 2} \dot{\phi} \left[1 + \eta \right] + \xi_l \eta + \Delta(\tau)  n^- + \Delta^*(\tau)  n^+ \right]\right\}.
\end{equation}
Note that in Eq.~(\ref{Zfull}) the explicit factorization of the terms into single-particle and paired/empty states is no longer necessary due to  ``locality''
of the ``dynamic partition function,'' $z_l$, after the Hubbard-Stratonovich decomposition. The contribution of the single-particle terms is simply given by the factor
of two in Eq.~(\ref{Zfull}).

To treat  the path integral (\ref{zk}), we note that it can be ``generated'' as a solution to the following differential equation for a ``density matrix,'' $\hat{\rho}_l$:
\begin{equation}
\label{rho}
{\partial \hat{\rho}_l \over \partial \tau} = - \hat{h}_l(\tau) \hat\rho_l \equiv
-\Biggl(
\begin{array}{cc}
\xi_l & \Delta(\tau) \\
\Delta^*(\tau) & -\xi_l
\end{array}
\Biggr)  \hat{\rho}_l, \mbox{ with } \hat{\rho}(0)= \hat{1}.
\end{equation}
The trace of the two-by-two ``density matrix'' evaluated at $\tau = \beta$ gives the desired partition function
\begin{equation}
\label{z=tr}
z_l = {\rm Tr}\, \hat{\rho}_l (\beta).
\end{equation}
This relation can be proven by writing a formal solution to Eq.~(\ref{rho}) as a $\tau$-ordered exponential and then expressing
it as a path integral to reproduce exactly (\ref{zk}).

To verify that the formulas obtained so far are consistent with what is known, let us consider the case of the classical mean-field,
where the order parameter is taken to be a constant $\Delta_{\rm BCS\, MF}(\tau) \equiv \overline{\Delta}_{\rm BCS}={\rm const}$. In this case the solution to Eq.~(\ref{rho})
is given by $\hat{\rho}_l^{(0)}(\beta) = \exp{\left(- \hat{h}_l \beta\right)}$. Since, $\hat{h}_l = \xi_l\, \hat{\tau}^z + {\rm Re}\,  \overline{\Delta}\, \hat{\tau}^x - {\rm Im}\,  \overline{\Delta}\, \hat{\tau}^y$, and we can write
\begin{equation}
\label{c+s}
\hat{\rho}_l^{(0)}(\beta) = \hat{1} \, \cosh\left( E_l \beta\right) - \left( {\bf n}_l \cdot \hat{\bm \tau} \right) \, \sinh \left(  E_l \beta\right),
\end{equation}
where $\hat{h}_l = E_l \left(  {\bf n}_l \cdot \hat{\bm \tau}  \right)$ with $\left| {\bf n}_l \right| = 1$, so that $E_l = \sqrt{\xi_l^2 + | \overline{\Delta}|^2}$ is the familiar quasiparticle spectrum in BCS theory, which in the pseudospin language translates into an effective magnetic field experienced by a pseudospin. Calculating the
trace, we recover the partition function of a spin-$1/2$ in a magnetic field of magnitude $\left| {\bf b}_l \right| = E_l$:
$z_l^{(0)} = 2 \cosh\left( E_l \beta\right)$. Now returning to Eq.~(\ref{Zfull}) and noticing
that $\left\{ 2 + 2 \cosh\left( E_l \beta\right) \right\} = \left\{ 2 \cosh\left({ E_l \beta \over 2}\right) \right\}^2$, we can write the  classical mean-field contribution to the partition function as follows:
\begin{equation}
\label{ZBCSMF}
Z_{\rm BCS\, MF} =
 \exp\Bigl[ - \beta \sum\limits_{l \in {\cal L}} \xi_l \Bigr]  \int {d^2  \overline{\Delta} \over \pi \tilde{g}} \exp\left\{ - {\beta \left|  \overline{\Delta} \right|^2  \over \tilde{g}} + 2 \sum\limits_{l \in {\cal L}} \ln \left[ 2 \cosh\left({ E_l \beta \over 2}\right) \right] \right\},
\end{equation}
where we recall that $\tilde{g} = g/(2 V_{\cal L})$. Varying the action with respect to $ \overline{\Delta}$, we indeed recover the familiar BCS self-consistency equation
\begin{equation}
\label{BCSSC}
{1 \over g}  = 2 V_{\cal L} \sum\limits_{l \in {\cal L}} {E_l^{-1}\,\tanh\left[ {E_l \beta \over 2} \right]}.
\end{equation}
We note that eventhough the classic BCS equation follows from the Richardson Hamiltonian, this zero-dimensional model does not have a true (classical) phase transition. In particular, if we calculate the  Riemann integral over $ \overline{\Delta}$  that appears within the classical mean-field approximation in Eq.~(\ref{ZBCSMF}), the resulting function $Z_{\rm BCS\,MF}(T)$ will be continuous in the vicinity of a nominal $T_{\rm c}$ ({\em e.g.}, one can expand the free energy into a Taylor series and obtain a zero-dimensional Landau theory, which leads to a continuous partition function expressed in terms of the error function; see Ref.~[\onlinecite{AVAI}] for details). If the underlying physical model is higher-dimensional, then a phase transition is anticipated,~\cite{AVAI} and we can interpret the temperature at which a derivative of the partition function over $T$  has the sharpest slope as a temperature where the phase transition occurs.  However, it is not only the partition function itself that is of primary interest, but also the trajectories that provide main contributions to it. In the weak-coupling limit, the transition point can in turn be identified (in the leading approximation with respect to $g$) with the point where a non-trivial solution to the self-consistency equation (\ref{BCSSC}) first appears, but in strong coupling this is not necessarily so.  We note that one can use the simple BCS result (\ref{ZBCSMF}) for estimates of $T_{\rm c} (g)$  by examining the partition function as explained above ({\em i.e.}, looking for a temperature where the slope of its second derivative is the sharpest). However, of course this procedure is not quantitatively reliable as it neglects superconducting fluctuations in real space (which are classical fluctuations for the purpose of determining $T_{\rm c}$), which have been excluded from Richardson model from the outset. In what follows, we will not address the very interesting question of determining  $T_{\rm c}$ in strong coupling, but instead will focus on the effects of quantum dynamics of the order parameter.

\section{Phase Fluctuations}
\label{sec:phase}

Let us now express the order parameter in Eq.~(\ref{Zfull}) explicitly as a product of a time-dependent amplitude part and a phase factor,
$\Delta(\tau) = \Delta_0 (\tau) e^{i\gamma(\tau)}$. To proceed further, we note that the first term (the factor of two) in the product in Eq.~(\ref{Zfull})
originates from single-particle states, which are free (real) spins and as such this factor of two is nothing but a partition function of a free spin-$1/2$. In the path-integral language, it can be ``generated'' by the action, which contains  just a Wess-Zumino term and no Hamiltonian. {\em I.e.}, we can use the following ``representation of the factor of two:''
\begin{equation}
\label{2}
2 = \int {\cal D}\, \Omega (\tau)e^{{i \over 2} \int\limits_0^\beta d\tau \dot{\phi} (1+\eta)}.
\end{equation}
Such Wess-Zumino terms appear in the factors $z_l$ in Eqs.~(\ref{Zfull}) and (\ref{zk}) as well and we get
\begin{eqnarray}
\label{Zfull_p}
\nonumber
Z_R =
e^{- \beta \sum\limits_{l \in {\cal L}} \xi_l } && \!\!\!\!
 \int {\cal D}^2 \left[ {\Delta(\tau) \over \pi \tilde{g}} \right]   \left[ \prod_{l \in {\cal L}} {\cal D}\,\Omega_l (\tau)  \right]
 \exp\left\{ -\int\limits_0^\beta d\tau \left[ {1\over \tilde{g}}  \Delta_0^2(\tau)  - {i \over 2} \sum\limits_{l \in {\cal L}}  \dot{\phi}_l (1+\eta_l) \right] \right\}\\
&& \!\!\! \times \prod_{l \in {\cal L}} \left\{ 1 + \exp{\left(-\int\limits_0^\beta d\tau \left[ \xi_l \eta_l - 2 \Delta_0 \cos\left( \phi_l - \gamma \right) \sqrt{1 - \eta_l^2} \right] \right)} \right\}.
\end{eqnarray}
We note here that the first term in the product, which is equal to one within our conventional Richardson model (\ref{CRH}), will have a more complicated form in the generalized Richardson Hamiltonian (\ref{GRH}), where it should be related to the partition function of an Ising model for real spins on singly-occupied sites.

We now perform the following change of variables (``gauge transformation'') $\phi_l \to \phi_l(\tau) + \gamma(\tau)$. The dependence of the action on the
overall phase of the order parameter  disappears from the last term in the product in Eq.~(\ref{Zfull_p}) and appears only in the Wess-Zumino term.
The corresponding $\gamma$-dependent part of the action therefore reads
\begin{equation}
\label{Sgamma}
S_\gamma =  -{i \over 2} \int\limits_0^\beta d\tau  \sum\limits_{l \in {\cal L}}  \dot{\gamma}(\tau) \left[1+\eta_l(\tau) \right].
\end{equation}
We can now evaluate the path integral over $\gamma(\tau)$, following Ref.~[\onlinecite{Zspin}] and keeping in mind the periodic boundary conditions for $\Delta(\tau)$ and ${\bf n} (\tau)$, so that $\gamma(\beta) - \gamma(0) = 2 \pi q$, with $q \in \mathbb{Z}$. Therefore, we obtain
\begin{equation}
\label{dgamma}
\int {\cal D}\gamma(\tau)\, e^{-S_\gamma} = \sum\limits_{q \in \mathbb{Z}} \exp\left\{ 2 i \pi q \sum_l \left[ 1 +  \eta_l(0) \right]/2 \right\}
\delta \left[ \sum\limits_l \dot{\eta}_l(\tau) \right].
\end{equation}
Hence, the phase fluctuations of the order parameter constrain the sum $ \sum_l \eta_l(\tau)$ to be equal to a constant at all times.
The resulting sum  over $q$ can be rewritten as an inverse discrete Fourier transform
$$
\sum\limits_{q \in \mathbb{Z}}  e^{2\pi i q x} = \sum_{{\cal N} \in \mathbb{Z}} \delta \left( x - {\cal N}  \right).
$$
Therefore, the result of path integration in (\ref{dgamma}) is
\begin{equation}
\label{Sgamma_res}
\int {\cal D}\gamma e^{-S_\gamma}  =  \sum_{{\cal N}  \in \mathbb{Z}}  \delta \left[ \sum\limits_{l \in {\cal L}} {1\over 2} \left\{ 1 + \eta_l(\tau) \right\} - {\cal N}  \right].
\end{equation}
The partition function for Anderson pseudospins  (\ref{Zfull_p}) reads 
\begin{eqnarray}
\label{Zfull_pp}
Z_R =
e^{- \beta \sum\limits_{l \in {\cal L}} \xi_l }
\sum\limits_{{\cal N}  =0}^\infty  \int {\cal D} \left[ {\Delta_0^2 \over 2 \pi \tilde{g}} \right]  \left[ \prod_{l \in {\cal L}} {\cal D} \, \Omega_l \right]
  \delta \left[  \sum\limits_{l \in {\cal L}} { 1 + \eta_l(\tau) \over 2} - {\cal N}  \right] e^{-S_{\Delta_0} - S_{\rm WZ} - S_{\rm eff}\left[ \{{\bf n}_l\}\right]},
\end{eqnarray}
where we limited the sum over ${\cal N} $ to positive values only because $(1 + \eta_l) \geq 0$ (if the set ${\cal L}$ is finite we can restrict  the sum to ${\cal N}  \leq V_{\cal L}$) and the path integral over the order parameter field includes only the dynamics of the modulus. In Eq.~(\ref{Zfull_pp}), $S_{\Delta_0}$  is the ``bare action'' for the order parameter field, $S_{\rm WZ}$ is the sum of all  Wess-Zumino terms for the pseudospins, and the interacting part of the effective action reads:
\begin{eqnarray}
\label{S2}
e^{- S_{\rm eff}\left[ \{{\bf n}_l\}\right]} = \prod_{l \in {\cal L}} \left\{ 1 + \exp{\left(-\int\limits_0^\beta d\tau \left[ \xi_l \eta_l - 2 \Delta_0 \cos\phi_l  \sqrt{1 - \eta_l^2} \right] \right)} \right\}.
\end{eqnarray}

We see that the effect of phase fluctuations of $\Delta(\tau)$ is to separate the partition function into ``sectors,'' where the total projection of the $z$-component of Anderson pseudospins is a constant integer at all times. This analogy can be made more explicit, if we imagine the associated real-time pseudospin dynamics, governed by the Bloch equation, $\dot{\bf M}_l = {\bf b}_l \times {\bf M}_l$, and where the  effective magnetic field is determined by ${\bf b}_l = \left( {\rm Re}\, \Delta_0(it), {\rm Im}\, \Delta_0(-it), \xi_l \right)$ [where $\Delta_0(-it)$ is the modulus of the order parameter properly analytically-continued to real times, $\tau \to it$]. The $\delta$-functions in Eq.~(\ref{Zfull_pp}) demand that the real-time dynamics of individual pseudospins must be correlated in such a way that they pin the  ``total pseudospin moment,'' $\sum_l M_l^z(t)$, to a constant. Note that these constraints  imposed by the phase fluctuations are in addition to the constraint that may be imposed by any mean-field treatment of the remaining path integral over the amplitude $\Delta_0$. From this, one can see that our ability or lack thereof to satisfy a certain mean-field (in a mesoscopic integrable system~\cite{R21,R22}), {\em e.g.} a constant amplitude such as in classic BCS mean-field, is determined by the initial conditions for the pseudospins.

  The physical meaning of all these results can be clarified if we first consider the subset of paired/empty states and recall that the density operator on a site, $l$, (\ref{density}) of the original model is given by $\hat{\rho}_l = 1 + \hat{\tau}_l^z$ for the paired/empty states  $l \in {\cal L}_2 \subset {\cal L}$, so that the Anderson ``spin-up'' corresponds to the existence of a Cooper pair and a ``spin-down'' to an empty site. Therefore the operator corresponding to the total number of Cooper pairs is given by
  \begin{equation}
  \label{CPtot}
\hat{\cal N}_{\rm CP} = {1 \over 2} \sum\limits_{l \in {\cal L}_2} \left[ 1 + \hat{\tau}_l^z \right]
\end{equation}
and the time-dependent field in the path-integral formalism corresponding to this operator is given by
  \begin{equation}
  \label{CPfield}
{\cal N}_{\rm CP}(\tau) = {1 \over 2} \sum\limits_{l \in {\cal L}_2} \left[ 1 + \cos{\theta_l(\tau)} \right] \equiv {1 \over 2} \sum\limits_{l \in {\cal L}_2} \left[ 1 + \eta_l(\tau) \right].
\end{equation}
From Eqs.~(\ref{Sgamma}) and (\ref{CPfield}), we see that the action that includes the phase of the order parameter can be written as follows
\begin{equation}
\label{Sgamma2}
S_\gamma[{\cal L}_2] =  -i  \int\limits_0^\beta d\tau  \dot{\gamma}(\tau) {\cal N}_{\rm CP}(\tau).
\end{equation}
{\em I.e.}, we recover the  fact that in the absence of gapless excitations, the phase of a Bose field operator and the number of bosons (Cooper pairs in our case) are canonically  conjugate operators, satisfying therefore the Heisenberg uncertainty principle. However, the effective action (\ref{S2}) may contain contributions from single-particle states as well [they are associated with the factor of one in the sum in Eq.~(\ref{S2})] and if we allow such states, {\em i.e.}, if $l \in {\cal L}_1 \ne \emptyset$, then the canonical conjugate to the phase, $\hat{\gamma}$, the way it is defined above,  will  also have a contribution from the real spins on singly-occupied sites. The meaning of the field $\left[1 + \eta_l(\tau)\right]/2$ is different for those singly-occupied states and relates to the $z$-component of the actual magnetic  moment of a site.  This suggests an interesting relation for the full phase action (\ref{Sgamma}), which now includes contributions from both paired/empty states and single-particle states:
\begin{equation}
\label{Sgamma3}
S_\gamma[{\cal L}] =  -i  \int\limits_0^\beta d\tau  \dot{\gamma}(\tau) \left[ {1 \over 2} {\cal N}_{\rm tot}(\tau) + S^z_{\rm tot}(\tau) \right],
\end{equation}
where ${\cal N}_{\rm tot}(\tau)$ and $S^z_{\rm tot}(\tau)$ are fields corresponding to the total number of particles and the total magnetic moment of the system. If single-particle states are completely gapped out as it is usually assumed, then all particles are bound in Cooper pairs, the total magnetic moment is identically zero and we recover the familiar conclusion summarized by Eq.~(\ref{Sgamma2}). But in general, the Hamiltonian version of Eq.~(\ref{Sgamma3}) will be a Heisenberg uncertainty relation/commutator, which involves both the superconducting  part (Anderson pseudospins) and a magnetic part  (real spins): $\left[ \hat{\gamma}, {1\over 2} \hat{\cal N}_{\rm tot} + \hat{S}^z_{\rm tot} \right] = i \hat{1}$.
We reiterate here that while our model is ``biased'' towards a superconducting state and has no magnetic interactions for real spins, a more general Richardson Hamiltonian [see, {\em e.g.}, Eq.~(\ref{GRH})] may have non-trivial magnetic interactions [see, {\em e.g.}, Eq.~(\ref{H1s})], which in principle may lead to a magnetic phase transition that would compete with superconductivity, {\em c.f.}, Refs.~[\onlinecite{ZhangSO(5)}], [\onlinecite{DemlerZhang}], [\onlinecite{MoonSS}], and [\onlinecite{VGSS}].

Both ${\cal N}_{\rm tot}$ and ${S}^z_{\rm tot}$ are certainly good quantum numbers and  are separately conserved. Hence, the phase, $\gamma$ fluctuates strongly (via the Heisenberg uncertainty principle) and since we treated these fluctuations exactly in Eq.~(\ref{Zfull_pp}),  the $\delta$-function constraints there  effectively enforce these underlying global conservation laws. An important question is whether we actually need to enforce them to describe a realistic superconductor. The classic description of an $s$-wave superconducting ground state requires no gapless excitations (${\cal L}_1 = \emptyset$) and hence the phase $\hat{\gamma}$ is identified with the phase of a Cooper-pair superfluid with  broken gauge symmetry (that is, $\gamma$ does not fluctuate). Per the same Heisenberg uncertainty principle, we must require then that either $\hat{\cal N}_{\rm CP} $ or $\hat{S}^z_{\rm tot} $ or both fluctuate strongly (in a closed system, it must be both, because the only way by which ${\cal N}_{\rm CP} $ can change  is by breaking Cooper pairs into single-particle excitations).

Another more technical way to argue in favor of the same conclusion is to consider a Richardson model or a more general (non-integrable) physical Hamiltonian from which it descends, weakly coupled to a bath and/or to a noisy magnetic field. Then, we are allowed to break weakly  some constraints associated with the global conservation laws. This can be done by ``softening'' the $\delta$-functions in Eq.~(\ref{Zfull_pp}): {\em E.g.}, we can represent each $\delta$-function as  a narrow Gaussian and then allow a finite width to the Gaussian, which would be equivalent to introducing a charging-energy-like term to the action $\delta S_\gamma \propto \int \dot{\gamma}^2 d\tau$  that penalizes phase fluctuations. Both these arguments suggest that to describe a realistic superconductor in the actual broken-symmetry phase, we have to suppress phase fluctuations, which can be accomplished by dropping the $S_\gamma$-term and the resulting constraints in the partition function (\ref{Zfull_pp}). This however brings up the question of whether the low-temperature state with broken gauge symmetry  will allow fluctuations of the  amplitude of the order parameter and if yes, whether they are purely mesoscopic or may involve more serious non-perturbative solutions.

\section{Amplitude Fluctuations}
\label{sec:amplitude}

\subsection{ Bogoliubov-de Gennes Equations in Imaginary Time}
\label{sec:BdG}

We now consider the amplitude fluctuations assuming that the phase fluctuations are suppressed. As it was shown in Sec.~\ref{sec:CSPI}, the partition function, originating from a non-trivial fluctuating order parameter field, is given by the trace of the density matrix $z_l[\Delta_0(\tau)] = {\rm Tr}\, \hat{\rho}(\beta)$ , which now is the solution to the following Bogoliubov-deGennes equation in imaginary time with a real, but generally time-dependent $\Delta_0(\tau)$,
\begin{equation}
\label{BdG}
{\partial \hat{\rho}_l \over \partial \tau} =  \hat{h}_l(\tau) \hat\rho_l \equiv
\Biggl(
\begin{array}{cc}
\xi_l & \Delta_0(\tau) \\
\Delta_0(\tau) & -\xi_l
\end{array}
\Biggr)  \hat{\rho}_l, \mbox{ with } \hat{\rho}(0)= \hat{1}.
\end{equation}
What is required at this stage is to find a general expression for $z_l[\Delta_0(\tau)] $ as a functional of the order parameter and to perform a variational analysis on the resulting effective action. This is equivalent to calculating the functional determinant in Eq.~(\ref{ZGHS11}), which appears within a more conventional treatment. This is a difficult  problem, which is intimately related to the problem of dynamics of a two-level system in a time-dependent magnetic field (generalized Landau-Zener problem), described via non-linear differential equations that have known analytic solutions only in a few special cases. While to determine the exact dynamics of pseudospins under an arbitrary perturbation, $\Delta_0(\tau)$, may not be possible, one can still get further insights by  taking advantage of the recent progress in understanding non-equilibrium BCS superconductivity~\cite{EY1} and the problem of dissipation due to externally driven two-level systems,~\cite{DG} where exact solutions can be obtained for a wide class of external perturbations associated with elliptic functions. Below, we explore solutions to Eq.~(\ref{BdG}) in some special cases  and generalize the results to express the functional determinant that arises within this class of dependencies in a compact form.

 However, let us start with a general analysis of the imaginary-time Bogoliubov-de Gennes equations (\ref{BdG}). Let us assume first that $\Delta_0(\tau) =
\Delta_0(-\tau)$, {\em i.e.} that it is an even function, which may occur ``naturally'' or via a periodic continuation from the physical imaginary-time interval, $\left[0,\,\beta\right]$  [all conclusions below can be generalized easily to the case where $\Delta_0(\tau) =
\Delta_0(2\tau_0-\tau)$]. Let us also consider a Nambu spinor $\chi(\tau) = {u(\tau)  \choose v(\tau)}$ and look for a solution to the following equations
\begin{eqnarray}
\label{BdG1}
\nonumber
&&{\partial_\tau u} = \xi u + \Delta_0(\tau) v;\\
&&{\partial_\tau v} = -\xi v + \Delta_0(\tau) u,
\end{eqnarray}
without specifying initial conditions. We also require that $\Delta_0(0) = \Delta_0(\beta)$, since it is a field that arises from a path integral in imaginary time. The corresponding function may have a ``natural'' period commensurate with $\beta$ or a single ``accidental'' period and in the latter case we shall periodically continue the function, $\Delta_0(\tau)$ defined on $\tau \in \left[ 0,\beta\right]$, such that it satisfies $\Delta_0(\tau) = \Delta_0(\tau + \beta),\, \forall \tau$.

Since we assumed that $\Delta_0(\tau) = \Delta_0(-\tau)$, the existence of a solution $\chi_1(\tau) ={u(\tau)  \choose v(\tau)}$ immediately implies that $\chi_2(\tau) = \left(-i\hat{\tau}^y \right) \chi_1(-\tau)\left(i\hat{\tau}^y \right) $ is also a linearly-independent solution. Hence a general solution to Eq.~(\ref{BdG1}) has the form
\begin{eqnarray}
\label{gen_sol}
\psi(\tau) = C_+ {u(\tau)  \choose v(\tau)} + C_- {-v(-\tau)  \choose u(-\tau)},
\end{eqnarray}
where $C_\pm$ are arbitrary constants determined by the initial conditions. Note that to determine the ``density matrix,'' $\hat{\rho}_l$, in Eq.~(\ref{BdG}), we need to find two solutions that satisfy the initial conditions, $\psi_1(\tau \to 0) = {1 \choose 0}$ and $\psi_2(\tau \to 0) = {0 \choose 1}$.
Let $u(0) =u_0$  and $v(0) = v_0$ be the initial conditions of a solution, $\chi_1(\tau)$, that we assume known. Then, from Eq.~(\ref{gen_sol}), we determine the solution that satisfies the first required initial  condition ({\em i.e.}, spin-up at $\tau = 0$) as follows
\begin{eqnarray}
\label{sol1}
\psi_1(\tau)= {1 \over u_0^2 + v_0^2} {u_0 u(\tau) + v_0 v(-\tau)  \choose u_0 v(\tau) -v_0 u(-\tau)}.
\end{eqnarray}
Per the same argument as above, the time-reversed to this solution, $\psi_2(\tau) =  \left(-i\hat{\tau}^y \right) \psi_1(-\tau)\left(i\hat{\tau}^y \right)$ satisfies the other initial condition ({\em i.e.}, spin-down at $\tau = 0$). Therefore, we conclude that if we know any solution to Eq.~(\ref{BdG1}), we can construct the $2 \times 2$ ``density matrix'' as follows:
\begin{equation}
\label{dm}
 \hat{\rho}(\tau) = \left\{ \psi_1(\tau); \left(-i\hat{\tau}^y \right) \psi_1(-\tau)\left(i\hat{\tau}^y \right) \right\},
  \end{equation}
where the solution, $\psi_1(\tau)$, and its time-reversed form the columns of $\hat{\rho}(\tau)$.  This results in the following ``partition function'' (functional determinant) of interest:
\begin{equation}
\label{fd}
z[\Delta_0(\tau)] ={u_0 \left[ u(\beta) + u(-\beta) \right] + v_0 \left[ v(\beta) + v(-\beta) \right]  \over u_0^2 + v_0^2}.
  \end{equation}
These results can be readily generalized to the case, where the order parameter is an even function with respect to an arbitrary $\tau_0 \in \left[0,\beta\right]$, {\em i.e.}, if $\Delta_0(\tau) = \Delta_0(2\tau_0 - \tau)$. The time reversal operation that generates another linearly independent solution can be written as follows
$\Psi_2(\tau,\tau_0) = \left(-i \hat{\tau}^y \right) \Psi_1(-\tau,-\tau_0) \left(i\hat{\tau}^y \right)$, where $\Psi_1(\tau,\tau_0)$ is a solution satisfying the initial condition $\Psi_1(0,\tau_0) \equiv {1 \choose 0}$ and, which itself can be constructed out of an arbitrary solution as follows $\psi(\tau) = C_+(\tau_0) {u(\tau-\tau_0)  \choose v(\tau-\tau_0)} + C_-(\tau_0) {-v(\tau_0-\tau)  \choose u(\tau_0-\tau)}$.

\subsection{Bogoliubov-de Gennes Equations and Supersymmetric Quantum Mechanics}
\label{sec:SUSY}

We see that if we know any particular solution to Eq.~(\ref{BdG1}) with arbitrary initial conditions, the problem of calculating the functional determinant is solved. However, it is of course the main challenge to find a particular solution. To shed light on the complexity of this general problem and to obtain a further interesting insight, we now take a detour  to point out a direct connection between the solvability of Bogoliubov-de Gennes equations (\ref{BdG1}) and supersymmetric Quantum Mechanics.

Let us introduce the following functions
\begin{equation}
\label{pRR}
p(\tau) = u(\tau) v(\tau),\,\, R_+(\tau) = {v(\tau) \over u(\tau)},\mbox{ and }\, R_-(\tau) = {u(\tau) \over v(\tau)}.
\end{equation}
From Eqs.~(\ref{BdG1}), we find
\begin{equation}
\label{Ric}
\left\{
\begin{array}{l}
\partial_\tau p(\tau)  = \Delta_0(\tau) p(\tau) \left[ R_+(\tau) +  R_-(\tau) \right];\\
\phantom{.}\\
\partial_\tau R_+(\tau)  = -2\xi R_+(\tau) + \Delta_0(\tau) \left[ 1 - R_+^2(\tau)  \right];\\
\phantom{.}\\
\partial_\tau R_-(\tau)  = 2\xi R_-(\tau)  + \Delta_0(\tau) \left[ 1 - R_-^2(\tau)  \right].
\end{array}
\right.
\end{equation}
The function, $p(\tau)$, is expressed in terms of the other two related functions $R_+(\tau) R_-(\tau) \equiv 1$:
\begin{equation}
\label{solp}
p(\tau) = p_0 \exp\left\{ \int\limits_0^\tau ds \Delta_0(s) \left[ R_+(s) +  R_-(s) \right] \right\},
\end{equation}
where $p_0$ is a constant of integration that can be set to one, $p_0 =1$, since we are looking for an arbitrary solution.
We see that Eqs.~(\ref{Ric}) for $R_\pm(\tau)$ are represented by a rather general Riccati equation, which has been studied for some 300 years
and which has known analytic solution only in a limited number of cases. But let us however proceed further and simplify the form of these equations
by introducing the following new variables:
\begin{equation}
\label{xWr}
x(\tau) = \int_0^\tau \Delta_0(s) ds,\,\, W(\tau) = {\xi \over \Delta_0(\tau)},\mbox{ and }\, r_\pm(\tau) = R_\pm(\tau) \pm W(\tau).
\end{equation}
We now assume also that $\Delta_0(\tau)$ does not change sign (which in fact is a requirement if phase fluctuations have been eliminated, since a change-of-sign in the order parameter should be incorporated into its phase dynamics). In this case, we can unambiguously determine $\tau(x)$ and treat all functions involved as
functions of $x$. We get
\begin{equation}
\label{Ric2}
r_\pm'(x) + r_\pm^2(x) = 1 + W^2(x) \pm W'(x) \equiv 1 + V_\pm(x).
\end{equation}
These are  Riccati-type equations as well, but they now have a form that is reminiscent to equations appearing in the context of supersymmetric Quantum Mechanics. To see the connection, we recall that a generic Riccati equation can always be reduced to the following form
\begin{equation}
\label{genRic}
f'(x) +{\phi'(x) \over \sigma(x)} f^2(x) =  {\sigma'(x) \over \phi(x)},
\end{equation}
such that a particular solution to Eq.~(\ref{genRic}) is written explicitly as $f_0(x) = {\sigma /\phi} + \phi^{-2} \left[{\rm const} + \int {\phi' \over \sigma \phi^2} dx \right]^{-1}$ and therefore the question of finding an analytic solution to a generic Riccati equation (\ref{genRic}) reduces to that of finding explicitly the functions $\phi(x)$ and $\sigma(x)$. In our case (\ref{Ric2}), we see that $\sigma_\pm(x) = \phi'_\pm(x)$, while the equations for $\phi_\pm(x)$ have the form
\begin{equation}
\label{SUSY}
\hat{H}_\pm\, \phi_\pm(x) = \left[ - {d^2 \over dx^2} + V_\pm(x) \right] \phi_\pm(x) = - \phi_\pm(x), \,\,\, x\in \left[0, L \right],
\end{equation}
where $L = \int\limits_0^\beta \Delta_0(s) ds$ is the period of the potentials, $V_\pm(x) = W^2(x) \pm W'(x)$, and $W(x)$ is defined in Eq.~(\ref{xWr}). We see that the operators $\hat{H}_\pm$ in the right-hand-side of Eq.~(\ref{SUSY}) are Schr{\"o}dinger operators associated with two superpotentials $V_\pm(x)$, which have the canonical form of those in  supersymmetric Quantum Mechanics~\cite{SUSYbook} and which in our case are determined by the underlying dynamics of the order parameter! Furthermore, since the $\Delta_0(\tau)$ has been periodically continued, Eqs.~(\ref{SUSY}) are actually Schr{\"o}dinger equations in a periodic superpotential.

Even though, for our purposes what is really needed is  the ``wave-function'' associated with just one (negative-energy) state, $E = -1$, we can easily examine whether the supersymmetric Schr{\"o}dinger equations admit zero modes (they do not). For this we can follow Ref.~[\onlinecite{PSUSY}] and notice that in a periodic potential the wave-functions are the Bloch-Floquet states, which for a zero-mode, if it were to exist, would translate into the condition $\phi_{0,\pm}(x+L) = e^{\pm \nu} \phi_{0,\pm}(x)$, with $\nu = \int_0^L W(x) =\beta \xi$. Since, the real factor $\nu$ is non-zero (apart from the state with $\xi =0$), there are no zero modes, the Witten index is zero, and therefore the supersymmetry is broken for our conventional $s$-wave superconductor. Admittedly, the significance of this fact is unclear (at least for topologically trivial superconductors studied here), but what however may be important is the fact that the existence of analytic solutions to the underlying Bogoliubov-de Gennes equations (\ref{BdG}) should be related to the existence of solvable supersymmetric potentials and {\em vice versa}. Another interesting approach that potentially may lead to progress would be to study quasiclassical solutions to Eqs.~(\ref{SUSY}), where the WKB method is known to work very well (it is exact in many notable cases). We shall however leave these questions for future work and explore below another means to treat the Bogoliubov-de Gennes equations to calculate the functional determinant of interest.

\subsection{Derivation of a Compact Expression for the Functional Determinant}
\label{sec:FDet}

\subsubsection{A Solvable Case with Non-Trivial Quantum Dynamics of the Order Parameter}
\label{sec:Solvable}

Up to this point, we have considered general properties of the Bogoliubov-de Gennes equations (\ref{BdG}) and found just one explicit solution corresponding to the ``trivial'' case of a constant order parameter, thereby recovering classic BCS theory in this language. For a further progress, it is desirable to examine the properties of some other exact solutions in less trivial cases, but as noted above the number of known solvable cases is quite limited. Fortunately, some additional insight comes from recent progress in the closely-related problem of non-equilibrium BCS superconductivity. There is a whole class of new solutions that have been recently obtained that not only admit exact analytic treatment of pseudospin real-time dynamics, but also amazingly satisfy the mean-field self-consistency constraint (for some specific real-time dynamics). Even though, as we shall see below, these solutions for $\Delta_0(\tau)$ are not at all optimal for minimizing the imaginary-time action in equilibrium, let us nevertheless examine some associated exact solutions for the density matrix. We will present below the simplest such solution, which is the imaginary-time version of the Ans{\"a}tz proposed by Levitov {\em et al.}~\cite{BLS} That is, let us seek the function, $R_+(\tau)$, [see, Eqs.~(\ref{pRR}) and (\ref{Ric})]  in the form
\begin{equation}
\label{Levit}
R_+(\tau) = 2 \xi f(\tau) - \dot{f}(\tau),
\end{equation}
where we, following Ref.~[\onlinecite{BLS}], identify $f(\tau) =\Delta_0^{-1}(\tau)$ so that the equation for $f$ reads
\begin{equation}
\label{EqfL}
\ddot{f} f = {\dot{f}}^2 - 1.
\end{equation}
This yields
\begin{equation}
\label{SolfL}
\Delta_0^{-1}(\tau) = f(\tau) = \omega^{-1} \cos{\left[\omega \left(\tau -\tau_0\right)\right]},
\end{equation}
where $\omega$ and $\tau_0$ are arbitrary constants for the purpose of satisfying Eq.~(\ref{EqfL}). However, we also have to
satisfy the periodicity requirement for the order parameter, $\Delta_0(0) = \Delta_0(\beta)$, which leads to two possibilities: (i)~If $\omega = 2 \pi n/\beta$, with
 $n \in \mathbb{Z}$, the solution~(\ref{SolfL}) is  ``naturally periodic'' with the period commensurate to $\beta$; (ii)~If $\omega \ne 2 \pi n/\beta$, but
 $\tau_0 = \beta/2$, it is an ``accidentally periodic'' solution. As we shall see below for the purpose of minimizing the imaginary-time action
the latter ``accidental'' periodicity is much preferable, while the naturally periodic solution is not even allowed in the case of (\ref{SolfL}).

In either case, the Ans{\"a}tz (\ref{Levit}) immediately leads to the following solution for $R_+(\tau) = v(\tau)/u(\tau)$
\begin{equation}
\label{SolR+L}
R_+(\tau) = {2\xi \over \omega} \cos{\left[ \omega \left(\tau -\tau_0\right) \right]} + \sin{\left[ \omega \left(\tau -\tau_0\right) \right]}
\end{equation}
and for the function $p(\tau) = u(\tau) v(\tau)$ [see, Eq.~(\ref{solp})]
\begin{equation}
\label{SolpL}
p(\tau) = {1 + {\omega /(2\xi)} \tan{\left[ \omega \left(\tau -\tau_0\right) \right]} \over \cos{\left[ \omega \left(\tau -\tau_0\right) \right]}} e^{2\xi(\tau-\tau_0)}.
\end{equation}
These solutions~(\ref{SolR+L}) and (\ref{SolpL}) together with Eqs.~(\ref{sol1}), (\ref{dm}), and (\ref{pRR}) determine the full ``density matrix'' ({\em i.e.}, the solution to the original equation (\ref{BdG}) as follows
\begin{eqnarray}
\label{SoldmL}
\nonumber
&&\hat{\rho}(\tau) =\left(
\begin{array}{cc}
\left\{\epsilon +  \tan(\omega \tau_0) \right\} \left\{\epsilon +  \tan{\left[ \omega \left(\tau -\tau_0\right) \right]} \right\};  &
 { \epsilon +  \tan\left[ \omega \left(\tau -\tau_0\right) \right] \over \cos(\omega \tau_0)} ; \\
{\epsilon + \tan (\omega \tau_0) \over \cos \left[ \omega \left(\tau -\tau_0\right) \right]}; & \cos^{-1} (\omega \tau_0)\cos^{-1} {\left[ \omega \left(\tau -\tau_0\right) \right]};
\end{array}
\right) {e^{\xi \tau} \over 1 +\epsilon^2}\\
&&+
\left(
\begin{array}{cc}
\cos^{-1} (\omega \tau_0)\cos^{-1} {\left[ \omega \left(\tau -\tau_0\right) \right]}; & {-\epsilon +  \tan(\omega \tau_0)  \over \cos\left[ \omega \left(\tau -\tau_0\right) \right]}; \\
{-\epsilon +  \tan{\left[ \omega \left(\tau -\tau_0\right) \right]}  \over \cos(\omega \tau_0)};& \left\{\epsilon -  \tan(\omega \tau_0) \right\} \left\{\epsilon -  \tan{\left[ \omega \left(\tau -\tau_0\right) \right]} \right\};
\end{array}
\right) {e^{-\xi \tau} \over 1 +\epsilon^2},
\end{eqnarray}
where we introduced $\epsilon = 2\xi/\omega$. One can explicitly verify that $\hat{\rho}(\tau)$ given by Eq.~(\ref{SoldmL}) above indeed satisfies Eq.~(\ref{BdG}) together with the initial condition $\hat{\rho}(0) =\hat{1}$.

The solution~(\ref{SoldmL}) looks complicated, but for the purpose of calculating the functional determinant (or equivalently the ``partition function,'' $z_l[\Delta_0(\tau)]$), we do not need its full form but need just its trace at $\tau = \beta$. Calculating this trace, we find an interesting result for this particular choice of the order parameter
\begin{equation}
\label{SolzL}
z_l\left[{\omega\over \cos  \left(\omega\tau -\omega\tau_0\right) }\right] = {\rm Tr}\, \hat{\rho}(\beta) = 2 \cosh{(\xi_l \tau)},
\end{equation}
which as we see does not depend on $\Delta_0(\tau)$ at all (neither on frequency nor on $\tau_0$) and is equivalent to a pseudospin not subject to any time-dependent $\Delta_0(\tau)$:~{\em I.~e.}, $z_l[\omega/\cos\left\{ \omega \left(\tau -\tau_0\right) \right\}]  = z_l[0]$. This is a very curious result indeed, because it suggests that the functional determinant may have a much simpler form than the actual ``density matrix'' used as a tool to calculate it.

To complete the analysis of this non-trivial fluctuation, let us examine the action evaluated for this particular ``trajectory'' of $\Delta_0(\tau)$, which has the form:
$S\left[\Delta_0(\tau)\right]  = {\omega^2  \over \tilde{g}} \int\limits_0^\beta {d\tau  \over \cos^2  \left(\omega\tau -\omega\tau_0\right) } -  \sum\limits_{l \in {\cal L}} z_l[0] $, with $z_l[0]$ given by (\ref{SolzL}). We notice that the first term diverges for the trajectories with ``natural periodicity'' because $\Delta_0(\tau)$ changes sign, which is not allowed if the phase fluctuations have been eliminated, but in any case such trajectories do not contribute to the partition function at all.   If however $\tau_0 = \beta/2$ and $\omega \beta < \pi$ (i.e., the order parameter is positive, $\forall \tau\,\in \left[0,\beta\right]$), we find immediately that the contribution to the action is
\begin{equation}
\label{SolSL}
S\left[{\omega\over \cos  \left(\omega\tau -\omega\beta/2\right) }\right] = {2 \omega \over \tilde{g}} \tan(\omega \beta) + S_{\rm FG} ,
\end{equation}
where the last term is the action of a non-interacting Fermi gas [{\em  c.f.}, Eq.~(\ref{SolzL})] and the first one is the energy cost to have a fluctuating order parameter in the form (\ref{SolfL}). Since there is no means to compensate this energy cost by adjusting the ``negative-energy'' term associated with $z_l$, we conclude that  the chosen ``trajectory'' of $\Delta_0(\tau)$ is a low-probability event in a low-temperature superconducting state. Note that since $\omega$ must be kept smaller than $\pi T$,  instantons of this type will completely die out in the ground state, but they may appear as  classical excitations at higher temperatures, including even the normal state.

\subsubsection{Functional Determinant on a Class of Elliptic Functions}
\label{sec:Ansatz}

Sec.~\ref{sec:Solvable} shows that while a tour-de-force derivation of the density matrix for a non-trivial fluctuating order parameter is not impossible, but generally is quite complicated, the actual result for the functional determinant may look very simple. To understand the origin of the ``mysterious'' simplification of the  complicated density matrix (\ref{SoldmL}) to the very simple-looking trace (\ref{SolzL}), we will rely on the recent work of Yuzbashyan~\cite{EY1} and a related work of Dzero and the author.~\cite{DG} 
%The details and a derivation of an extended family of explicit solutions will be published under a different cover,~\cite{DG} and below we  present some general considerations [we note here that even though the discussion below may appear mathematically formal, it is not intended to be or even appear to be a proof by any means, but it is rather a set of arguments put together in an attempt to unify a set of results observed for certain types of exactly solvable quantum dynamics of $\Delta(\tau)$]:

Let us consider a particular quantum trajectory of the order parameter $\Delta_0(\tau)$ and analytically continue this function from $\tau \in \mathbb{R}$ to complex values $z = \tau + i t \in \mathbb{C}$. One can formulate a sensible general framework in terms of a $z$-dependent ${\cal S}$-matrix, $\hat{\cal S}(z)$, but we will consider here only its analytically continued form on the real-time axis (which  is equivalent  to a Feynman-Wick rotation at $T=0$). Let us now use  this analytical continuation to relate the Bogoliubov-de Gennes equation (\ref{BdG}) for the density matrix in imaginary time to the corresponding Schr{\"o}dinger equation for the $S$-matrix in real time $t$:
 \begin{equation}
\label{Sm}
i{\partial \hat{S}_l \over \partial t} =  \hat{h}_l(it) \hat{S}_l(t) =
\Biggl(
\begin{array}{cc}
\xi_l & \Delta_0(it) \\
{\Delta_0^*(it)} & -\xi_l
\end{array}
\Biggr)  \hat{S}_l(t), \mbox{ with } \hat{S}_l(0)= \hat{1}.
\end{equation}
where $\Delta_0(it)$ and $\hat{h}_l(it)$ are symbolic notations for the dynamic order parameter and the Hamiltonian properly analytically-continued to
 real times, correspondingly. The Hamiltonian, $\hat{h}_l (it) ={\rm Re}\, \Delta_0(it) \hat{\tau}^x + {\rm Im}\, \Delta_0(it) \hat{\tau}^y + \xi_l \hat{\tau}^z  = \left( {\bf b}_l \cdot \hat{\bm \tau} \right)/2$, belongs to the two-dimensional representation of the Lie algebra, $\mathfrak{su}(2)$, while the unitary $S$-matrix belongs to the two-dimensional representation of the  $SU(2)$ group. Note that there is no need to specify the dimensionality of the matrix representation of operators in Eq.~(\ref{Sm}), which can be viewed as an equation of motion in the abstract group, {\em i.e.} $\hat{h}_l (it) \in \mathfrak{su}(2) \sim \mathfrak{so}(3)$ and $\hat{S}(t) \in SU(2), \forall t$.

We can also write an associated Schr{\"o}dinger equation for  spinor wave-function, $\Psi ={\psi_\uparrow \choose \psi_\downarrow}$
  \begin{equation}
\label{SE}
i\dot{\Psi} =
\Biggl(
\begin{array}{cc}
\xi_l & \Delta_0(it) \\
{\Delta_0^*(it)} & -\xi_l
\end{array}
\Biggr)  \Psi,
\end{equation}
Just like in Sec.~\ref{sec:BdG}, we can argue that if we know a particular solution to Eq.~(\ref{SE}) that satisfies an arbitrary initial condition,
we can always construct another linearly independent solution with the help of a time-reversal operation that now reads $\left( -i \hat{\tau}^y \right) \Psi^*(t) \left( i \hat{\tau}^y \right)$ and hence a full $S$-matrix can be constructed using one particular solution to (\ref{SE}). Let us note here that a key motivation for studying the analytically-continued form of the Bogoliubov-de Gennes Eqs.~(\ref{BdG}) and (\ref{BdG1}) is  that  if we know the solution to Eq.~(\ref{Sm}) or just a particular solution to Eq.~(\ref{SE}), which has the familiar form of a Schr{\"o}dinger equation, we should be able to analytically continue the result back to imaginary time [so that in some sense $\hat{S}(-i\tau) \to \hat{\rho}(\tau)$] and therefore calculate the functional determinant.

Now, let us narrow down a class of functions considered from that of arbitrary order parameters (analytically continued from the values $\tau \in \left[ 0, \beta \right]$ to complex arguments), $\Delta_0(z)$,  to those that are periodic along the $it$-axis, {\em i.e.},  $\Delta_0(z) = \Delta_0(z + i\beta_t)$,
where $\beta_t \in \mathbb{R}$ is the corresponding period. Let us also assume that the original fluctuation  is a meromorphic and periodic function  along the $\tau$-axis as well, {\em i.e.}, $\Delta_0(z) = \Delta_0(z + \beta_\tau)$, which is either due to a natural periodicity with the period commensurate to the inverse temperature $\beta_\tau = \beta/n$ or some other period unrelated to the fact that the relation $\Delta_0(0) = \Delta_0(\beta)$ must hold (see, {\em e.g.}, previous Sec.~\ref{sec:Solvable}, where depending on the parameters both cases can be realized). This narrower class of functions represents elliptic functions,~\cite{Elbook} with primitive periods $\beta_\tau$ and $i \beta_t$, which are arbitrary constants at this point.
%Our choice of this (very large) class of functions is of course not accidental and relates to the following: (i)~We actually know exact solutions to Eq.~(\ref{SE}) for some specific elliptic functions. (ii)~The results of a systematic analysis of the non-equilibrium BCS problem by Yuzbashyan {\em et al.} strongly suggest to look for a saddle-point  of the corresponding action in the form of an elliptic function or its limit.
%
%Even though, the discussion of this section has so far  been rather formal, we now will proceed with general intuitive arguments that led us to the main conjecture for the functional determinant below. These arguments are supported by the observations in all special cases where we know exact solutions, but we do not attempt any formal proof and do not systematically look for counter-examples to see if it is reasonable to extend those conclusions to a wider class of functions. So, the level of mathematical rigor of the arguments below is admittedly not very high.

If $\Delta_0(z)$ is an elliptic function with the primitive periods $(\beta_\tau,i\beta_t)$ as defined above, then the Schr{\"o}dinger equation (\ref{Sm})
is that describing a spin-$1/2$ under a periodic-in-time perturbation and so let us look for a solution of (\ref{Sm}) in a Bloch-Floquet-type  form,
$\hat{S}(t) = \hat{S}_{\rm p}(t) e^{i{\cal E}t} + \hat{S}^\dagger_{\rm p}(t) e^{-i{\cal E}t},$
 where $\hat{S}_p(t) = \hat{S}_p(t + \beta_t)$ is a periodic $2\times2$-matrix and ${\cal E}$ is a constant. On the other hand, we could have used the same Floquet argument for the original Bogoliubov-de Gennes equation to argue that the ``density matrix'' may be written in a similar form, $\hat{\rho}(\tau) = \hat{\rho}_{\rm  p,+} (\tau) e^{{\cal E}\tau} +  \hat{\rho}_{\rm  p,-} (\tau) e^{-{\cal E}\tau}$, where now $\hat{\rho}_{\rm  p, \pm}(t) = \hat{\rho}_{\rm p,\pm} (t + \beta_\tau)$ is a ``periodic part of the density matrix'' [{\em c.f.}, Eq.~(\ref{SoldmL})], and ${\cal E}$ is the same as before.  These arguments suggest themselves to be generalized in
the form of a solution $\hat{\cal S}(z) = \hat{\cal S}_{\rm  p,+}(z) e^{{\cal E}z} + \hat{\cal S}_{\rm  p,-}(z) e^{-{\cal E}z}$, where $z = \tau + it$ and $\hat{\cal S}_{\rm  p,\pm}(z)$ is an ``elliptic matrix function'' of a complex argument, $z \in \mathbb{C}$, such that 
$\hat{\cal S}_{\rm  p,\pm}(z) = \hat{\cal S}_{\rm  p,\pm}(z + n \beta_\tau + im \beta_t ),\, \forall n,m \in \mathbb{Z}$.

If we now assume that the periodicity of $\Delta_0(z)$ along the $\tau$-axis is commensurate with the ``natural'' period, $\beta$, then we find immediately that
$ \hat{\cal S}_{\rm  p,\pm}(z) = 1/2$ because of the initial condition $\hat{\cal S}(0) = {\hat \rho}(0) = \hat{1}$ and the imposed periodicity and therefore we conclude that
\begin{equation}
\label{FD1}
z[\Delta_0(\tau)] = 2 \cosh{\left({\cal  E}   \beta \right)},
\end{equation}
where the current assumptions are that $\Delta_0(z)$ is an elliptic function with the  periods $(\beta_\tau,i\beta_t)$ such that $\beta/\beta_\tau \in \mathbb{Z}$ and
$\beta_t$ is arbitrary and the solution of the Schr{\"o}dinger equation (\ref{Sm}) has a Bloch-Floquet form.

To get a more useful expression for the ``partition function,'' $z$, let us now focus on a dependence $\Delta_0(it)$ slow enough so that there are no level
crossings taking place. Consider now a pseudospin, described by the spinor $\Psi_{\rm l} (t)$ of Eq.~(\ref{SE}), evolving from the initial state that is an eigenstate of the Hamiltonian at $t=0$, {\em e.g.}, we can take it to represent to a pseudospin moment opposite to the ``initial magnetic field,'' ${\bf b}_l(0) =  \left( \Delta_0(0),  0 , \xi\right)$, which lies in the $XZ$-plane. The adiabaticity assumption immediately tells us that the quantum-mechanical phase ``collected'' after the completion of  a single cycle,  $t:\,0\,\to \beta_t$ of the magnetic field, ${\bf b}(t)$, is given by the expression in the exponential below
 \begin{equation}
\label{phases}
\Psi_{\rm l} (\beta_t) = \Psi_{\rm l} (0) e^{-i\left(\gamma_{\rm Berry}  + \gamma_{\rm dyn}\right)},
\end{equation}
where $\gamma_{\rm Berry}$ is the Berry phase, which is determined by the following flux through the area, ${\cal A}_{\bf b}$, swept by ${\bf b}(t)$ over
the cycle, $t:\,0\,\to \beta_t$,
  \begin{equation}
\label{Bphase}
\gamma_{\rm Berry}= {\xi \over 2} \int\limits_{{\cal A}_{\bf b}}
{d {\rm Re}\, \Delta_0(it) \wedge  d {\rm Im}\, \Delta_0(it) \over \left[ \xi^2 + \left| \Delta_0(it) \right|^2 \right]^{3/2}},
\end{equation}
 and $\gamma_{\rm dyn}$ is the dynamical phase given  by
 \begin{equation}
\label{dynphase}
\gamma_{\rm dyn}=  \int\limits_0^{\beta_t} \sqrt{ \xi^2 + \left| \Delta_0(it) \right|^2} dt,
\end{equation}
where the integrand can be easily recognized as an ``instantaneous eigenenergy'' of the corresponding spin Hamiltonian (which in turn represents
the energy of an excitation in a superconductor subject to such a fluctuation). Note that we could have taken the other initial condition corresponding
to a pseudospin pointing along the ``initial magnetic field,'' which would have evolved into a state with the dynamical phase, which is a complex conjugate to $(i \gamma_{\rm dyn})$ above. We can recall now that since we are interested in the $S$-matrix modulo its periodic part, we can construct the remainder
out of the two phase factors, which therefore gives exactly the desired $\left( {\cal E} \beta_t \right)$ that appears in Eq.~(\ref{FD1}).

If we now make a further simplifying assumption and consider a  fluctuation, $\Delta_0(\tau -\tau_0)$, which is described by an even function of its argument (see, previous Sec.~\ref{sec:Solvable}) such that the analytically-continued $\Delta_0(it)$ is also real-valued, we immediately find that the effective ``magnetic field'' simplifies to ${\bf b}_l(t) =  \left( \Delta_0(it),  0 , \xi\right)$. Therefore, the area swept by any such dependence in the parameter space is  zero and the Berry phase (\ref{Bphase})  vanishes identically as well. Note that this conclusion would also hold if we assume that the order parameter is an odd function of $(\tau-\tau_0)$, such that it leads to a purely-imaginary $\Delta_0(it)$ (let us recall that phase fluctuations in imaginary time have been eliminated). Under these assumptions, we can  identify the factor arising from the ``non-periodic'' part of the $S$-matrix/``density matrix'' with the dynamical phase, $\left( {\cal  E}   \beta_t\right) = \gamma_{\rm dyn}$ to obtain the following result for the functional determinant:
\begin{equation}
\label{conjecture}
z_l[\Delta_0(\tau)] = 2 \cosh{\left[ {\beta \over \beta_t}   \int\limits_0^{\beta_t} \sqrt{ \xi_l^2 + \left| \Delta_0(it) \right|^2} dt  \right]},
\end{equation}
where we have restored the index $l$ that parameterizes the sites of the original Richardson model.

\subsubsection{The Adiabaticity Requirement}
\label{sec:Adiab}

Eq.~(\ref{conjecture}) has appeared after a chain of rather general arguments, which however included a number of additional assumptions. Let us reiterate these assumptions: We have assumed that $\Delta_0(\tau)$ can be analytically-continued from $\tau \in \left[0,\,\beta\right] \in \mathbb{R}$ to the complex plane, $\mathbb{C}$, and that the resulting function $\Delta_0(z)$ is an elliptic function with two primitive periods $(\beta_\tau,i\beta_t)$ such that $\beta/\beta_\tau \in \mathbb{Z}$. We also have further assumed that the there exists $\tau_0 \in \mathbb{R}$ such that the order parameter is either an odd or an even function of
the argument $\Delta(\tau -\tau_0)$, which ensures that the Berry phase vanishes. Finally, we have assumed that the time-dependence of the analytically-continued order parameter, $\Delta_0(it)$ is ``slow enough,'' such that no level crossings take place.

The last assumption is the most restrictive and one may wonder about the accuracy and domain of applicability of the conjecture (\ref{conjecture}) and in particular about the meaning of ``slow enough'' in the adiabaticity assumption. This question has been addressed by Gangopadhyay, Dzero, and the author in Ref.~[\onlinecite{DG}] in the context of two-level-system dynamics in superconducting qubits. Mathematically,  Ref.~[\onlinecite{DG}]  presented an extended class of exact solutions associated with elliptic functions describing the driving field (which represent a generalization of the anomalous solitons discussed in the amazing paper of Yuzbashyan in Ref.~[\onlinecite{EY1}]). Here we reiterate only key facts relevant to our paper:

The following functional dependencies of $\Delta_0(it)$ admit exact explicit solutions of the associated Eqs.~(\ref{SE}) and (\ref{Sm}):
\begin{equation}
\label{Dzero}
\Delta_0({\bf e}, it) = \Omega_a + \Omega_+ {1 - \eta_+ {\rm sn}\,^2 \left( \omega t, \kappa \right) \over 1 + \eta_- {\rm sn}\,^2 \left( \omega t, \kappa \right)},
\end{equation}
where ${\bf e} = \left(e_1, e_2, e_3 \right)$ is a shorthand to describe three parameters that appear in the following equation [{\em c.f.}, Eq.~(\ref{EqfL})], $\dot{f}^2 = (f-e_1)(f-e_2)(f-e_3)$, and are subject to the constraint $e_1 + e_2 + e_3 = 0$. A solution to the equation for $f$ above can be expressed in terms of the Weierstrass elliptic function, which is related to the order parameter (\ref{Dzero}). The other constants in Eq.~(\ref{Dzero}): $\Omega_a$, $\Omega_+$, $\omega$, and $\kappa$, are not free but are determined uniquely by ${\bf e}$~\cite{DG} [{\em E.g.}, $\kappa = (e_2 - e_3)/(e_1-e_3)$]. Finally, the function ${\rm sn}\, \left( \cdot, \kappa \right)$ in Eq.~(\ref{Dzero}) is the doubly-periodic Jacobian elliptic function.

The derivation of the associated solution involves Bloch representation of Anderson pseudospins as follows ${\bf M}(t) = {1 \over 2} \Psi^\dagger(t) \hat{\bm \tau} \Psi(t)$, where of course ${\bf M}^2(t)\equiv 1/4$ so that ${\bf M}(t) \in S^2$, which is the standard Bloch sphere.  The equations of motions for the Bloch vectors follow from the Schr{\"o}dinger equation in a standard way and yield the familiar Bloch equations:
\begin{equation}
\label{BS}
\dot{\bf M}(t) = -i  \Psi^\dagger(t) \left[  \hat{\bm \tau}, \hat{h}_l(it) \right]\Psi(t) ={\bf b}_l(t)
\times {\bf M}(t)
\end{equation}
where the ``magnetic field'' in the cross product is exactly as in Sec.~\ref{sec:Ansatz}: ${\bf b}_l(t) =  \left(  \Delta_0(it),  0 , \xi_l\right)$ with $\Delta_0(it)$
given by Eq.~(\ref{Dzero}). Therefore,  the equations of motion (\ref{SE})  in the $SU(2)$-group have been reduced to equations of motion (\ref{BS}) on a sphere, $S^2$. Let us recall that  $S^2 = SU(2)/U(1)$, therefore Eq.~(\ref{BS}) has less direct information than the original Schr{\"o}dinger equation. It turns out that the ``missing part'' is  exactly the sought-after overall time-dependent $U(1)$ phase of the wave-function, which we expect to reduce to the sum of the Berry phase and a dynamic phase discussed in the previous Sec.~\ref{sec:Ansatz} [in the case of dependence (\ref{Dzero}), the Berry phase is zero].

Using  the Ans{\"a}tz proposed by Yuzbashyan in Ref.~[\onlinecite{EY1}], one can find~\cite{DG} the solutions, ${\bf M}(t)$,  to Eqs.~(\ref{BS}), expressed in terms of elliptic functions with the same periodicities that the elliptic function  (\ref{Dzero}) in accordance with the suggested generalization of the Floquet argument to elliptic functions, as discussed in the previous Section~\ref{sec:Ansatz}. Using these exact solutions one can construct the full $S$-matrix describing the motion in $SU(2)$. It can be done by parameterizing the components of the  spinor in Eq.~(\ref{SE}) as follows $\psi_{\uparrow/\downarrow}(t) = {1 \over \sqrt{2}} \left|a_{\uparrow/\downarrow}(t)\right|\, e^{i\gamma(t) \mp {i \over 2} \theta(t)}$. One can see that while the amplitudes and relative phase are directly related to the ``instantaneous'' direction of the Bloch vector $\left|a_{\uparrow/\downarrow}(t)\right| = \sqrt{1 \pm 2 M^z(t)}$ and $\theta(t) = \arctan\left[ {M^y(t) / M^x(t)} \right]$, the common phase $\gamma(t)$ depends on the trajectory in a non-local way and to determine it, one has to go back to the  Schr{\"o}dinger equation~(\ref{SE}). This indeed can be done and the phase, $\gamma$, can be found (this part has to be done numerically for generic parameters).
The main conclusion of this analysis is that if $\Omega_a$ is small, this exact phase is essentially indistinguishable from the dynamic phase described by Eq.~(\ref{dynphase}) [however, for any non-zero $\Omega_a$, Eq.~(\ref{conjecture}) is not exact]. As $\Omega_a$ increases up to a critical value, $\Omega_a^{\rm (cr)}$, level crossings start taking place and  the absolute value of the quantal phase is suppressed compared to the adiabatic result. In all cases  considered, the adiabatic quantal phase is either equal or larger than the exact phase, and therefore it can be viewed as an estimate from above. Ref.~[\onlinecite{DG}] also indicates that for all $\Omega_a < \Omega_a^{\rm (cr)}$, the adiabaticity condition is satisfied and in this case the compact (\ref{conjecture}) expression for the functional determinant can be used.

\subsubsection{Explanation of the ``Fermi Gas'' Result Obtained in the Solvable Case of Sec.~\ref{sec:Solvable}.}
\label{sec:Expl0}

This discussion was initially motivated by the ``paradox'' found in the fully-solvable case described in Sec.~\ref{sec:Solvable}. We remind that the contribution
to the action of  a pseudospin moving in the presence of a non-trivial fluctuating $\Delta_0(\tau)$ described by Eq.~(\ref{SolfL}) turned out to be completely independent of the parameters of this fluctuation and was found to be identical to the corresponding contribution expected in a Fermi gas, i.e., in the absence of
any order parameter whatsoever, $\Delta(\tau) \equiv 0$. Another part of the ``paradox'' was that the full solution for the ``density matrix,'' $\hat{\rho}(\tau)$, was very cumbersome (\ref{SoldmL}) and the simplification occurred at the final stage of calculating its trace, which led us to Eq.~(\ref{SolzL}) for $ {\rm Tr}\, \hat{\rho}_l(\beta) = 2 \cosh{(\xi_l \tau)}$.

These paradoxes can be now resolved with the help of Eqs.~(\ref{conjecture}) and (\ref{Dzero}). One can consider various limits of the function (\ref{Dzero}), in particular, that of $\Omega_a \to 0$, which leads to the following expression of the dynamic order parameter considered previously by Yuzbashyan~\cite{EY1} and Levitov {\em et al.}:~\cite{BLS}
 \begin{equation}
 \label{dn}
 \Delta_0({\bf e},it)\Bigr|_{\Omega_a \to 0} = \omega\,\, {\rm dn} \left[ \omega t, {2 \sqrt{\kappa} \over 1 +\kappa} \right].
 \end{equation}
On the other hand, the limit $\kappa \to 1$ leads to ${\rm sn}\, (u,1) = \tanh{u}$ and Eq.~(\ref{Dzero}) reproduces the anomalous soliton of Ref.~[\onlinecite{EY1}]. If we now take both limits, {\em i.e.}, $\kappa \to 1$ and $\Omega_a \to 0$, we find
 \begin{equation}
 \label{dn_lim}
 \Delta_0({\bf e},it)\Bigr|_{\Omega_a \to 0; \kappa \to 1} = {\omega \over \cosh(\omega t)}.
 \end{equation}
An analytical continuation of this function to imaginary time yields $ \Delta_0({\bf e},\tau)\Bigr|_{\Omega_a \to 0; \kappa \to 1} = \omega/ \cos{(\omega t)}$,
which is {\em exactly} the soliton studied in Sec.~\ref{sec:Solvable}. Note that this soliton is {\em not} an elliptic function, but a circular function, because it has only one incommensurate period in the $\tau$-``direction.'' However, it is a limiting case of a proper elliptic function, with its period $\beta_t$ taken to infinity. Therefore, per the arguments of Sec.~\ref{sec:Ansatz} and using Eq.~(\ref{conjecture}), we are to write the ``partition function'' as follows: 
\begin{equation}
\label{demystery}
z_l[\omega/ \cos{(\omega t)}] = 2 \cosh{\left[ \lim_{\beta_t \to \infty} {\beta \over \beta_t}   \int\limits_0^{\beta_t} \sqrt{ \xi_l^2 + {\omega^2 \over \cosh^2{(\omega t)}}} dt  \right]} = 2 \cosh\left( \beta \xi_l \right).
\end{equation}
Now, it is easy to see  the origin of the paradoxical result (\ref{SolzL}): Since  $\cosh^{-2}(\omega t)$ decays exponentially  with increasing $t$, the soliton term in Eq.~(\ref{demystery}) vanishes in the $\left({\beta_t \to \infty}\right)$-limit and does not contribute to the integral. We therefore recover the correct result (\ref{SolzL})!

Now that the origin of the result (\ref{SolzL}) in the exactly-solvable case is understood, we can use the exact solution to get another insight into the range of
applicability of Eq.~(\ref{conjecture}). The list of assumptions for the validity of Eq.~(\ref{conjecture}) includes that $\beta$ be a natural rather than an accidental period of $\Delta_0(\tau)$ (enforcing this periodicity via a periodic repetition of $\Delta_0(\tau)$ from $\tau \in \left[0, \beta \right] \to \mathbb{R}$ would not necessarily work, because the resulting periodically-continued function may not have the desired analytical properties and any arguments based on them would become unreliable). Hence, we do not expect Eq.~(\ref{conjecture}) to work in the accidentally-periodic cases, but the
exactly-solvable example (\ref{SolfL}) with $\tau_0 =\beta/2$ and $\forall \omega \in  \mathbb{R}$ shows that at least in this particular case Eq.~(\ref{conjecture}) does work correctly. It remains unclear at this stage, whether this result is an artefact of the particular dependence (\ref{SolfL})  or it is rather an indication that Eq.~(\ref{conjecture}) applies to a wider class of elliptic functions and their limits with accidental $\beta$-periodicity.

\subsection{Contribution of Elliptic Trajectories to the Partition Function}
\label{sec:Seff}

Now let us summarize our findings (conjectures) and present the following expression for the contribution to the partition function of those specific elliptic trajectories,  $\Delta_0(\tau) \to \Delta_0(z) \in {\rm Ell}$, for which our expression for the functional determinant applies:
\begin{eqnarray}
\label{ZEll}
&& \delta Z_{\rm Ell} = e^{- \beta \sum\limits_{l \in {\cal L}} \xi_l }
 \int\limits_{\Delta_0(z) \in {\rm Ell}} {\cal D} \left[ {\Delta_0^2(\tau) \over 2 \pi \tilde{g}} \right] e^{-S\left[ \Delta(\tau) \right]},
 \mbox{   where }\\
&& S\left[ \Delta(\tau) \right] \gtrsim {1 \over \tilde{g}} \int\limits_0^\beta \left|  {\Delta_0(\tau)} \right|^2 
- 2 \sum\limits_{l \in {\cal L}} \ln \left[ \cosh \left( {\beta \over 2\beta_t}   \int\limits_0^{\beta_t} \sqrt{ \xi_l^2 + \left| \Delta_0(it) \right|^2} dt    + {1\over 2} \gamma_{\rm Berry} \left[ \Delta_0(it) \right] \right) \right],
\nonumber
\end{eqnarray}
where just as before $\Delta_0(it)$ corresponds to an analytically-continued order parameter, which leads to an elliptic function with the primitive periods $(\beta_\tau,i\beta_t)$ or a limit of such an elliptic function. In Eq.~(\ref{ZEll}), we have also included the Berry phase contribution, which in general should be present, but whenever $ \Delta_0(it)$ is either purely real or purely imaginary, the Berry phase vanishes identically. 

We have already verified that the action in Eq.~(\ref{ZEll}) reproduces  the exact results in  certain exactly-solvable cases. Note here that the classic BCS result (\ref{ZBCSMF}) is certainly reproduced exactly as well, because a constant function represents a trivial elliptic function, $\Delta_0(\tau) \equiv \Delta_0(it) \equiv \overline{\Delta}= {\rm const}$ and so we can use Eq.~(\ref{ZEll}), and after an integration recover the correct ``partition function'' of a spin in a constant magnetic field, $z_l^{(0)} = 2 \cosh\left( E_l \beta\right)$, which corresponds to the BCS mean-field. One can also argue based on (\ref{ZEll}) that the classical mean-field is indeed a true minimum on the space of these elliptic functions, ${\rm Ell}$. Let us consider an order parameter $\Delta_0(\tau) = \overline{\Delta} + \delta\Delta(\tau)$, where $\delta\Delta(\tau)$ is in some sense small. Let us expand the action in Eq.~(\ref{ZEll}) assuming that $\delta\Delta(\tau)$ does not induce a Berry phase. Hence a correction to the relevant part of the action is (we consider the low-temperature limit, $\beta \to \infty$):
\begin{equation}
\label{dS2}
\delta S_2 = -2 \overline{\Delta}_0 \sum\limits_{l \in {\cal L}}  {\beta \over \beta_t} \int\limits_0^{\beta_t} {\delta\Delta(it) \over \sqrt{ \xi_l^2 + \overline{\Delta}_0^2}} dt
\end{equation}
Note now that the integral above can be equivalently written as an average of $\delta\Delta(it)$ over an infinite number of periods, ${\beta \over   \beta_t} \int\limits_0^{\beta_t} \delta\Delta(it) dt \equiv \lim\limits_{n \to \infty} {\beta \over n \beta_t} \int\limits_0^{n \beta_t} \delta\Delta(it) dt$. Therefore, the contour of integration over $z = \tau +it$ is that going from $0$ to $i \infty$. The difference between this integral and that going along the  $\tau$-axis is
$\left[\int\limits_0^{i \infty} \delta\Delta(it) dt - \int\limits_0^{\beta \to \infty} \delta\Delta(\tau) d\tau \right] = (-2 \pi i) \sum_{z_i} {\rm res}\, \delta\Delta(z_i) = 0$, {\em i.e.},
it is given by the sum of all residues of $\delta\Delta(z)$ enclosed in the first quadrant. It is equal to zero, per one of the elementary properties of elliptic functions,~\cite{Elbook} which states that the sum of all the residues of an elliptic function inside a period-parallelogram always vanishes. Therefore, one can write the first variation  of the both parts of the action in terms of the same function, $\delta\Delta(\tau)$: $\delta S = 2  \overline{\Delta}_0 \left[ (1/\tilde{g}) -   \overline{\Delta}_0 \sum\limits_{l \in {\cal L}} \left( \xi_l^2 + \overline{\Delta}_0^2 \right)^{-1/2} \right]    \int\limits_0^{\beta\to \infty} d\tau \delta\Delta(\tau) = 0$, which is satisfied for the BCS mean-field.

One may wonder, if one can use the analytical properties of elliptic functions to bring the integral that appears in Eq.~(\ref{ZEll}) to the $\tau$-axis in a similar way, {\em  i.e.},  to the form $\to  \int\limits_0^\beta \sqrt{ \xi_l^2 + \left| \Delta_0(\tau) \right|^2} $?  We know however that this substitution can not generally be correct, because some non-trivial solutions that we have analyzed manifestly contradict  this assumption. However, we have proven above that for all relevant  fluctuations in the immediate ``vicinity'' of the classic BCS mean-field the substitution above would work. One can explicitly verify that the interesting property of this (generally incorrect) substitution  is that the variational analysis of the functional, $\tilde{S}\left[f_0(\tau)\right] =  \int\limits_0^\beta d\tau {1 \over \tilde{g}} \left|  {f_0(\tau)} \right|^2  - 2 \sum\limits_{l \in {\cal L}} \ln \left[ \cosh \left( {1 \over 2}   \int\limits_0^{\beta} \sqrt{ \xi_l^2 + \left| f_0(\tau) \right|^2} d\tau \right) \right] $ ({\em i.e.}, the constraint ${\delta \tilde{S} \over \delta f_0} = 0$)  indeed immediately selects the classical mean-field $f_0(\tau) \equiv \overline{\Delta}_{\rm BCS} ={\rm const}$ as the {\em only saddle point}. Hence, one can use the expression for $\tilde{S}$ above to determine the contributions to the partition function due to Gaussian quantum fluctuations in the vicinity of the BCS mean-field (for simplicity, we consider the low-temperature limit only). The result is not unexpected and is quite ``boring,'' taking the following form for the usual BCS superconductor ({\em i.e.}, the parameter space, ${\cal L}$, is momentum space):
\begin{equation}
\label{nearBCS}
Z_{\rm near\, BCS} =e^{-{{\cal F}_{\rm BCS} \over T}}\, \int {\cal D} \left[ {\delta\Delta(\tau) \over \pi \tilde{g}} \right] e^{ - {9 V \over 4 g} \int\limits_0^\beta \delta\Delta^2(\tau) d\tau},
\end{equation}
where ${\cal F}_{\rm BCS}$ is the energy of the classical mean-field BCS state, $T =\beta^{-1}$, and $V$ is the actual physical volume and hence the contribution of these mesoscopic fluctuations to observables in a bulk system is negligible.

It is alluring to attempt a variational analysis of the action in Eq.~(\ref{ZEll}) to see if there could exist other saddle points apart from the classical mean-field. However, the variational analysis would  be problematic, because the action in Eq.~(\ref{ZEll}) contains ``apples and oranges,'' that is two functionals of different functions, $\Delta_0(\tau)$ and $\Delta_0(it)$, which are related  in a  non-trivial way via an analytical continuation.  However, to make the case that non-linear soliton contributions are important (we should distinguish here between instantons, which are trajectories that connect classical minima,~\cite{Polyakov} and fictitious at this stage new minima, which we dub solitons), one does not necessarily need to find true quantum minima, finding any quantum trajectory that corresponds to the energy smaller than mean-field would suffice. To clarify the content of this (open) problem, let us introduce two parameters $\Delta_1^2 = {\beta^{-1} \int\limits_0^\beta \Delta_0^2(\tau) d\tau}$ and $\Delta_2$ such that $\left({\beta \over 2} \sqrt{ \xi_l^2 + \Delta_2^2} \right)= \left({\beta \over 2\beta_t}   \int\limits_0^{\beta_t} \sqrt{ \xi_l^2 + \left| \Delta_0(it) \right|^2} dt     + {1\over 2} \gamma_{\rm Berry} \left[ \Delta_0(it) \right] \right)$. In these notations, the action in Eq.~(\ref{ZEll}) takes the form [{\em c.f.}, Eq.~(\ref{ZBCSMF})]:
\begin{equation}
\label{D1D2}
S[\Delta_1,\Delta_2] = {\beta \Delta_1^2 \over \tilde{g}} - 2 \sum\limits_{l \in {\cal L}} \ln \left[ 2 \cosh\left({ \sqrt{ \xi_l^2 + \Delta_2^2}  \beta \over 2}
\right)\right].
\end{equation}
Since the first term is always positive and the second one is always negative, we are to look for ways to minimize $\Delta_1$ and maximize $\Delta_2$. In the classical BCS mean-field $\Delta_1 = \Delta_2$  and there is no room for any additional variation, but in the functional~(\ref{ZEll}) such additional
variations are in principle allowed. One can check that the analytical continuation of solitons of the non-equilibrium BCS problem with natural periodicity do not produce a ``good'' solution at least at $T=0$. If on the other hand, we take Eq.~(\ref{ZEll}) as a given functional and consider various trial functions without attempting to prove that they actually satisfy the formal domain of validity of the Ans{\"a}tz, we immediately find a variety of dependencies that do ``better than classical mean-field'' in term of energetics. However, these ``results'' should be taken with a grain of salt, because there is no way to determine the actual range of validity of (\ref{ZEll}) beyond those dependencies associated with known integrable spin dynamics and the most natural explanation for any accidental solution obtained within a trial-and-error analysis of (\ref{ZEll}) is that it is probably beyond the applicability of the method.  On the other hand,  there appears to exist no proof that such solitons are impossible. Looking at the rich structure of the functional determinant, it appears conceivable that there exist trajectories in the huge functional space spanned by, $\Delta(\tau)$, that do not just  collapse into the mesoscopic term (\ref{nearBCS}), but instead provide more noticeable contributions to the action. A numerical analysis of some non-linear solutions, guided by the analytical result~(\ref{ZEll}), will be published elsewhere.

\section{Summary}
\label{sec:Summary}

This paper presents an analysis of non-perturbative fluctuation phenomena in the pairing model.
The key step of this analysis is a decomposition of the partition function of the Richardson model into spin and
pseudospin terms. It is shown that such factorization is possible for a generalized Richardson
model that includes both BCS and spin interactions. Even though we have not presented here a
theory to describe both types of non-trivial interactions on an equal footing, the development of such an extension
is straightforward~\cite{G} and would lead to a two-order-parameter theory expressed in terms of two ``global''
Hubbard-Stratonovich fields.~\cite{ZhangSO(5)} The analysis of phase fluctuations presented here indicates that these
interactions will be competing  and that such competition can be enforced via a commutation
relation between  the density and spin density and the overall phase. However, the present paper  has focused on the analysis
of a simpler canonical Richardson model that has no magnetic interactions. Even though the spin sector of this Richardson model
is trivial, the existence of this (single-particle) sector is important for the possible existence of any non-trivial fluctuations
of the amplitude of the order parameter in the low-temperature phase.

The main technical part of the paper involves a calculation of functional determinants that appear in the
non-linear effective action expressed in terms of the Hubbard-Stratonovich field. We have shown that the  Anderson pseudospin
language and in particular its coherent-state path-integral representation lead to  practically useful and physically
intuitive insights into the structure of the functional determinants for non-trivial quantum trajectories. Therefore, this approach
may be much preferable to the conventional Grassmann path integral method. We have shown that a functional determinant
is given by the trace of a density matrix that satisfies the Bogoliubov-de~Gennes equations in imaginary time. This leads
to a differential equation of the Riccati type, which  is directly related to the supersymmetric Schr{\"o}dinger equation
with superpotentials determined by the imaginary-time dynamics of the order parameter, $\Delta(\tau)$. Let us note here
that a particularly promising direction for further research could be to use the WKB-method to treat the relevant differential equations.

In Secs.~\ref{sec:Ansatz} and \ref{sec:Seff}, we proposed an explicit, compact expression for the functional determinant for
a certain large class of elliptic functions and the arguments that led us to the conjecture (\ref{ZEll}) involved an analytical continuation
of the Bogoliubov-de~Gennes equations in imaginary time to the real-time axis (or more generally to the complex plane, $z = \tau +it$),
such that the problem could be mapped onto that of a two-level system in a time-dependent magnetic field determined by quantum dynamics.
This is a known, very complicated problem, but we have taken advantage of some recent exact results and our recent work
on an extension of these results to analyze a family of exact solutions that are associated with elliptic functions. These results have led us to Eq.~(\ref{ZEll}), which provides a useful intuition for the effective action of the model and suggests that the functional determinant, that  is often treated as a thing-in-itself, can actually be calculated and is related in a very straightforward way to the dynamical and Berry phases of a pseudospin moving in a ``magnetic field,'' determined by the quantum dynamics of a fluctuation. Let us reiterate however that a formal justification of our solution applies only to adiabatic dependencies on the specific class of elliptic functions with the periods along  the imaginary and real-time axes. %Let us note here
%that even if the exact solutions we used were not available, the choice of the functional sub-space of elliptic functions would have been quite natural
%in this problem. Indeed, any constraint  imposed by the self-consistency equation on the order parameter
%should translate into a constraint on collective dynamics of pseudospins (with $\Delta(it)$ representing a very large classical pseudospin
%moment). So at some (very qualitative) level the problem should be related to that of magnetic moments precessing around
%one another.

An important open question is whether the considerations presented in this paper can be generalized to other types of functions, $\Delta(\tau)$,
which are not associated with any elliptic functions that lead to integrable  pseudospin dynamics. A particularly promising avenue here could be to use the reverse-engineering approach for constructing exact solutions as described in Refs.~[\onlinecite{DG}] and [\onlinecite{BWZW}], which effectively implies a change-of-variables from the Hubbard-Stratonovich field, $\Delta(\tau)$, to the generators, ${\bm \Phi}(t)$, which govern the dynamics of the $S$-matrix, $\hat{S}(t)= \exp \left[ - {i \over 2} {\bm \Phi}(t) \cdot \hat{\bm \sigma} \right]$, satisfying the proper Bogoliubov-de~Gennes equations. It would also be interesting to see whether a chaotic rather than integrable dynamics~\cite{DYA} can be realized under any circumstances in this model. Quite generally such dynamics, if at all possible, are not expected to lead to energetically favorable contributions to the action,  because the ``trivial'' term that gives an energy penalty to any non-zero order parameter configuration corresponds to the average of $|\Delta|^2$, while the second non-trivial term that favors superconductivity  contains contributions from different sites, and if the dynamics exhibit a ``chaotic behavior'' in the parameter space, ${\cal L}$, the signs in the second term would  fluctuate strongly from site to site and are expected to average out to zero instead of lowering the corresponding energy. This argument supports the approach to use regular elliptic trajectories that describe a synchronized collective behavior of the pseudospins. Another open question relates to the role of the Berry phase in the   functional determinant~(\ref{ZEll}). All exact solutions we have analyzed [that are sensible to describe thermodynamics, where the constraint $\Delta(0) = \Delta(\beta)$ must be imposed] have a trivial (zero) pseudospin Berry phase. This however represents a limitation in our ability to solve Eqs.~(\ref{BdG}), rather than an indication that Berry phase terms are unimportant.

Finally, we reiterate the main question posed in this paper and the arguments of the last Sec.~\ref{sec:Seff}, which suggest that non-perturbative soliton trajectories that co-exist with the classical mean-field are not impossible  and in fact the rich general structure of the functional determinant suggests that the construction of such quantum fluctuations may be possible at least in some modification of the model (which may  involve interactions for real spins). Generally, the right question to ask would be whether there exists any fermion model that exhibits breaking of continuous symmetry and  such that its low-temperature phase allows non-perturbative soliton solutions for a component  of the Hubbard-Stratonovich field that is normally considered ``massive?'' In other words, can the non-linear effective action for the Hubbard-Stratonovich field develop any other minima  apart from the classical mean-field? A proof that no such solutions exist would confirm the fundamentals of classical spontaneous symmetry breaking and mathematically would imply that there is no need to study complicated non-linear actions at $T=0$ [such as the non-linear  effective action in Eq.~(\ref{Seff})] and that in an infinite system they should crossover to a functional delta-function of the type, $e^{-S[\Delta(\tau)]} \propto \delta \left[ \left| \Delta(\tau) \right| - \overline{\Delta}_{\rm MF} \right]$, {\em c.f.}, Eq.~(\ref{nearBCS}). On the other hand, even a single example of an order-parameter trajectory that is energetically beneficial to the classical mean-field would seriously question this fundamental conjecture. We know that any such trajectory, if at all possible, can not be anywhere near classical mean-field (in the functional space of allowed fluctuations), but the possibility of a non-perturbative solution not adiabatically-connected to the mean-field has certainly not been ruled out.

{\em Acknowledgements:}~The author is grateful Vladimir Gritsev for discussions and hospitality at the University of Fribourg and to the  members of the Maryland condensed matter theory group for a number of illuminating discussions, specifically to Maxim Dzero, Anirban Gangopadhyay, Tigran Sedrakyan, and Justin Wilson. This work was supported by the Department of Energy.

\bibliography{QF}

\end{document}